\documentclass[preprint, preprintnumbers,showpacs, showkeys, aps, prd, amsmath, amsfonts, amssymb,
   eqsecnum, nofootinbib, floatfix, secnumarabic]{revtex4-1}

\usepackage{epsf}
\usepackage{epsfig}
\usepackage{graphics}
\usepackage{graphicx}
\usepackage{amssymb}

\def\lsim{\mathrel{\raise.3ex\hbox{$<$\kern-.75em\lower1ex\hbox{$\sim$}}}}
\def\gsim{\mathrel{\raise.3ex\hbox{$>$\kern-.75em\lower1ex\hbox{$\sim$}}}}

\begin{document}
\preprint{CUMQ/HEP 188}

\title{ \Large  Dark Matter in the Higgs Triplet Model }

\author{Sahar Bahrami$^1$\footnote{Email: sahar.bahrami@concordia.ca}}
\author{Mariana Frank$^1$\footnote{Email: mariana.frank@concordia.ca}}

\affiliation{ $^1 $Department of Physics,  
Concordia University, 7141 Sherbrooke St. West ,
Montreal, Quebec, Canada H4B 1R6.}

\date{\today}

\begin{abstract}

The inability to predict neutrino masses and the existence of the dark matter are two essential shortcomings of the Standard Model. The Higgs Triplet Model provides an elegant resolution of neutrino masses via the seesaw mechanism. We show here that  introducing vectorlike leptons in the model also provides a resolution to the problem of dark matter. We investigate constraints, including the invisible decay width of the Higgs boson and the electroweak precision variables, and impose restrictions on model parameters. We analyze the effect of the  relic density constraint on the mass  and  Yukawa coupling of   dark matter. We also calculate the cross sections for indirect and direct dark matter detection and  show our model predictions for  the neutrino and muon fluxes from the Sun, and the restrictions they impose on the parameter space.  With the addition of vectorlike leptons, the model is completely consistent with dark matter constraints, in addition to improving electroweak precision and doubly charged mass restrictions, which are rendered consistent with present experimental data.

\end{abstract}
\maketitle



 \section{Introduction}
  \label{sec:intro}
  
The LHC discovery of the Higgs Boson \cite{LHCHiggs} with properties consistent with that of the Standard Model (SM) Higgs, while providing a spectacular experimental confirmation of the SM, continues to raise questions about SM completeness and about scenarios responsible for  new physics beyond the SM. In addition, non-collider experimental results confront the SM with two major puzzles: neutrino masses and  the existence  of dark matter. 

The phenomenon of neutrino oscillations
shows that at least two neutrinos have nonzero but small masses, located around sub-eV
scale \cite{pdg}. The fact that the neutrino flavor structure is so different from that of
quarks and leptons is a puzzle and may indicate that neutrinos are Majorana particles. Many models have been proposed to explain tiny neutrino masses. The see-saw mechanism, in
which right-handed neutrinos are introduced with large
Majorana masses \cite{seesaw, seesaw2}, is  perhaps the simplest way to explain tiny neutrino masses.  The most direct way for implementation of this mechanism for generating
neutrino masses is to enlarge the particle content of the SM by a complex triplet scalar field, yielding the so-called Higgs Triplet Model (HTM)
{\cite{htm, htm2}. The neutrino mass problem is resolved at the cost of introducing only this additional Higgs representation, together with its associated vacuum expectation value (VEV), but without extending the symmetry of the model. The  triplet scalar field also plays a role in in leptogenesis \cite{Hambye:2003ka}.


At the same time, evidence from astrophysics and cosmology indicate that the ordinary
baryonic matter is not dominant in the Universe. Rather, about 25\% 
of energy density of the Universe is comprised of a  non-luminous and non-absorbing matter, called
dark matter (DM). While current observations  indicate that most of the matter in the Universe
is non-baryonic dark matter,  they do not provide information on what this dark
matter consists of. Since the Standard Model (SM), which has been extremely
successful in describing all current collider data, does not contain any dark
matter candidates, a great deal of effort has gone into providing viable candidates, or alternatives scenarios (models which include a DM candidate naturally).  The latter type of models do so at the expense of extra symmetries and a much enriched particle content. For models lacking natural candidates, a common method is 
  to consider the simplest
additions to the SM that can account for the dark matter. In these models, the
SM particle content is extended by a small number of fields, and a new discrete
symmetry is introduced to guarantee the stability of the dark matter
particle. Several variations can be obtained depending on the number and type
of new fields (e.g. a scalar, fermion, or vector, a singlet or a doublet under SU(2),
etc.) and on the discrete symmetry imposed ($Z_2$, $Z_3$, $\ldots$). 
 
 In this work, we look at the Higgs Triplet Model for a resolution to both neutrino masses and dark matter problems.  The resolution to neutrino masses, alluded to in the above,  is well-known \cite{htm}. The complex triplet couples to left-handed leptons, yielding Majorana masses for the neutrinos through $L=2$ lepton flavor violating terms \cite{htm2}, while  also contributing to  type-II leptogenesis \cite{Ma:1998dx}. In addition, extra degrees of freedom that couple to the SM Higgs at the tree-level insure cancellation of quadratic divergences to the Higgs mass \cite{Chakraborty:2014xqa}, a mechanism where scalars are favored. Additional support for the model comes from the observation that  heavy particles with strong couplings to the Higgs field can strengthen the electroweak phase transition, through the entropy release mechanism from both bosons and fermions \cite{Chang:2014aaa}.   
 
Unfortunately, as it stands, the Higgs Triplet Model lacks a dark matter candidate. Resolutions to this problems were proposed: some with additional Higgs triplets,  where the neutral component of the additional (real) Higgs representation can act as a DM candidate \cite{FileviezPerez:2008bj}, another where an {\it additional} $SU(2)_L$ triplet scalar fields with hypercharge $Y=1$ is added \cite{Kanemura:2012rj}. In this work, we investigate the possibility that the DM candidate is provided through  the introduction of a complete fourth-generation of vectorlike leptons, comprised of $SU(2)_L$ doublets  plus charged and neutral $SU(2)_L$ singlets \cite{Joglekar:2012vc}. A simpler extension of the SM with only one fourth generation vectorlike lepton doublet  coupling to a triplet Higgs field, which gives Majorana mass to a pseudo-Dirac fourth neutrino has been considered in \cite{Arina:2012aj}. 

Vectorlike pairs of fermions, unlike their chiral counterparts, are able to have mass explicitly through the gauge-invariant bilinear interaction in the Lagrangian $M_f f^\dagger f$. There is no reason why such pairs of vectorlike fermions do not exist, and many theories such as string theories and D-brane theories, often give rise generically to vectorlike states~\cite{Dijkstra:2004cc}.  Since the mass of the vectorlike fermions are not generated through the Yukawa couplings, the loop contributions involving the Higgs decouple faster than for chiral fermions. Thus the constraints from the current Higgs data, precision electroweak observables and direct searches are less severe for vectorlike fermions than for chiral fermions. 

Originally, there has been a great deal of interest in vectorlike leptons as a resolution to preliminary data indicating an
  enhanced Higgs decay rate to diphotons, while the Higgs production cross section was in agreement with expectations from SM. The diphoton rate is increased
through loops of mixed vectorlike leptons. A vectorlike doublet and a vectorlike singlet
allow for both Yukawa couplings and Dirac masses. The resulting mixing leads to a sign flip of the
coupling of the lightest lepton to the Higgs field, yielding constructive interference with the
standard model (SM) amplitude for $h \to \gamma \gamma$. 

This does not have to be so in the Higgs Triplet Model, where contributions from vectorlike leptons can be offset by contributions from charged and doubly charged Higgs bosons. However new effects of vectorlike leptons can arise.  Previous analyses have shown that their presence affects the mass bounds and decay patterns of the doubly charged Higgs boson \cite{Bahrami:2014ska}, improving consistency with present experimental data.

We extend our previous considerations in \cite{Bahrami:2014ska} to explore the possibility that, introducing a new parity symmetry making all new vectorlike leptons odd, and prohibiting the mixing with the ordinary SM leptons,  the lightest particle which is
odd under this symmetry (a singlet neutrino) becomes stable on cosmological timescales, and could have properties consistent with it being a candidate for the dark matter of the universe. Note that in a simple heavy fourth generation extension of SM, the heavy neutrino does not qualify as a dark matter
due to its rapid  annihilation to SM particles via $Z$ boson exchange \cite{Keung:2011zc}. Leptonic dark matter candidates with unsuppressed couplings to the $Z$ boson, such as
ordinary fourth generation neutrinos are also excluded by limits from direct detection \cite{Angle:2008we}. 
This constraint can be relaxed in the model considered here, as  the two singlet neutrinos in the model have no couplings, or very small couplings, to the $Z$ boson. 

Suppression of the lightest neutrino couplings to the $Z$ boson can also evade present experimental limits from LEP on masses of new charged and neutral particles \cite{pdg}. Measurements of the $Z$ boson width restrict the number of active neutrinos to three, which further restrict the mass of the new neutrino to $M_N>39$ GeV for a Majorana, and $M_N>45$ GeV for a Dirac neutrino, precluding the viability of a neutrino which couples to the $Z$ boson as a candidate for light dark matter. While, as we will show,  we can relax these constraints here, the new states will have an effect on the precision electroweak parameters, which we calculate and use to restrict the parameter space. We then analyze the consequences of the model by requiring consistency with the invisible Higgs width and non-collider experimental data, particularly with direct and indirect dark matter searches.   The  relic density, an indication of the abundance of dark matter in the early universe, as measured by PLANCK satellite \cite{Ade:2013zuv}, is one of the most stringent constraints on any model of DM, as well as direct detection experiments search for  spin-independent (SI) or spin-dependent
(SD) interactions with target nuclei, which can be detected by nuclear recoil experiments.
Indirect detection experiments searches looking for gamma ray excesses measure the annihilation products of DM, and their predictions must also be  tested in a model of DM. Finally, ultra-high energy neutrino experiments measure the neutrino flux and flavor composition at astrophysical sources. We analyze the predictions for all of these in our model and indicate the constraints on vectorlike neutrino mass and coupling which restrict our parameter space. 

Our work is organized as follows. In the next section, Sec. \ref{sec:model}, we summarize the basics features of the Higgs Triplet Model with vectorlike leptons. We proceed by examining the electroweak precision constraints  in the HTM in Sec. \ref{sec:STU}, where we present a numerical analysis on  restrictions coming from the oblique parameters on the masses of the doubly charged Higgs bosons and relevant Yukawa coupling. We discuss the invisible decay width of the Higgs boson  in Sec. \ref{sec:IDW}. Then in  Sec. \ref{sec:RD}, we calculate the dark matter relic density and indicate the restrictions it imposes on  the mass of the dark matter and on the Yukawa couplings. These restrictions are then applied to the evaluation of the spin-dependent and spin-independent in the direct detection of dark matter in Sec. \ref{sec:DD}, and of the annihilation cross section of dark matter in Sec. \ref{sec:ID}. We discuss detection of DM at colliders in Sec. \ref{sec:LHC}, and then investigate the fluxes of muons and neutrinos from the Sun in  Sec. \ref{sec:NMF}.  We summarize our findings and conclude in Sec. \ref{sec:conclusion}. 
\section{The Higgs Triplet Model with Vectorlike Leptons}
\label{sec:model}
Here we review briefly the HTM with vectorlike leptons, a more detailed version which has appeared in our previous work \cite{Bahrami:2014ska}. The symmetry group of the HTM is the same as that of the SM, with the particle content enriched by {\it (a)}  the addition of one triplet scalar field $\Delta$ with hypercharge $Y=1$, and with vacuum expectation value (VEV) $v_\Delta$:
\begin{eqnarray}
\Delta =
\left[
\begin{array}{cc}
\frac{\Delta^+}{\sqrt{2}} & \Delta^{++}\\
\frac{1}{\sqrt{2}}(\delta+v_\Delta+i\eta) & -\frac{\Delta^+}{\sqrt{2}} 
\end{array}\right],
\end{eqnarray}
and {\it (b)} a vectorlike fourth generation of leptons\footnote{We assume vectorlike quarks to be heavy \cite{pdg} and decouple from the spectrum.}, to include:   one $SU(2)_L$ left-handed lepton doublet $L_L^\prime=(\nu^\prime_L, e^\prime_L)$, right-handed charged and neutral lepton singlets, $\nu_R^\prime$ and $e_R^\prime$,  and the mirror right-handed lepton doublet, $L_R^{\prime\prime}=(\nu^{\prime\prime}_R, e^{\prime\prime}_R)$ and left-handed charged and neutral lepton singlets $\nu_L^{\prime\prime}$ and $e_L^{\prime\prime}$, as listed in Table \ref{tab:VLrepresentations}. Note that $v_\Delta$ is restricted to be small by the see-saw mechanism, which requires generation of small neutrino masses, and by the $\rho$ parameter. In general we can  assume, conservatively, $v_\Delta \lsim 5$ GeV \cite{Kanemura:2012rs}. 

\begin{table}[htbp]
\caption{\label{tab:VLrepresentations}\sl\small Representations of vectorlike leptons, together with their quantum numbers under $SU(3)_C \times SU(2)_L \times U(1)_Y$.}
  \begin{center}
 \small
 \begin{tabular*}{0.99\textwidth}{@{\extracolsep{\fill}} c| ccccccc}
 \hline\hline
	Name &${\cal L^\prime}_L$ &${\cal L^{\prime \prime}}_R$ &${e^\prime}_R$ &${e^{\prime \prime}}_L$ &${\cal \nu^\prime}_R$ &${\cal \nu^{\prime \prime}}_L$ 
	\\
  \hline
  Quantum Number &$(\mathbf{1}, \mathbf{2}, -1/2)$ &$(\mathbf{1}, \mathbf{2}, -1/2)$ &$(\mathbf{1}, \mathbf{1}, -1)$ &$(\mathbf{1}, \mathbf{1}, -1)$ &$(\mathbf{1}, \mathbf{1}, 0)$ &$(\mathbf{1}, \mathbf{1}, 0)$  \\
      \hline
    \hline
   \end{tabular*}
\end{center}
 \end{table}
The Lagrangian density for this model contains, in addition to the SM terms, kinetic, Yukawa for ordinary leptons, explicit terms for the vectorlike leptons, and potential terms:
\begin{eqnarray}
\mathcal{L}_{\rm{HTM}}= \mathcal{L}_{\rm{kin}}+\mathcal{L}_{Y}+\mathcal{L}_{\rm VL}-V(\Phi,\Delta), 
\end{eqnarray}
where
\begin{eqnarray}
\mathcal{L}_Y&=&-\left[\bar{L}_L^ih_e^{ij}\Phi e_R^j+\rm{h.c.}\right] -\left[ h_{ij}\overline{L_L^{ic}}i\tau_2\Delta L_L^j
+\rm{h.c.} \right],~~~~ \label{nu_yukawa}
\end{eqnarray}
are the Yukawa interaction terms for the ordinary leptons, with  $h^{ij}_{e}$ a  3$\times 3$ complex matrix, and $h_{ij}$ a $3\times 3$ complex symmetric Yukawa matrix.  
Additionally, with the vectorlike family of leptons as defined above, 
\begin{eqnarray}
\mathcal{L}_{\rm VL}&=&-\big [M_{L}{\bar L}_L^\prime L_R^{\prime \prime}+M_{E}{\bar e}_R^{\prime} e_L^{\prime \prime}+ M_{\nu}{\bar \nu}_R^{ \prime} \nu_L^{\prime \prime}+\frac12 M_\nu^\prime {\overline \nu^{\prime c}_R} \nu_R^\prime + \frac12 M_\nu^{\prime \prime} {\overline \nu^{\prime \prime c}_L} \nu_L^{\prime \prime}+h_E^\prime ({\bar L}_L^\prime \Phi )e_R^\prime
 \nonumber \\
&& +h_E^{\prime \prime} ({\bar L}_R^{\prime \prime} \Phi )e_L^{\prime \prime}+h_\nu^\prime ({\bar L}_L^\prime \tau \Phi^\dagger )\nu_R^\prime +h_\nu^{\prime \prime} ({\bar L}_R^{\prime \prime} \tau \Phi^\dagger )\nu_L^{\prime \prime} + h^\prime_{ij}\overline{L_L^{\prime\, c}}i\tau_2\Delta L_L^{\prime\, }
+h^{\prime \prime}_{ij}\overline{L_R^{\prime \prime \, c}}i\tau_2\Delta L_R^{\prime\prime } \nonumber \\
&&+  \lambda_E^i ({\bar L}_L^\prime \Phi )e_R^i +  \lambda_L^i ({\bar L}_L^i \Phi )e_R^\prime + \lambda^\prime_{ij}\overline{L_L^{ ic}}i\tau_2\Delta L_L^{\prime}
+ \lambda^{\prime \prime}_{ij}\overline{L_R^{ ic}}i\tau_2\Delta L_R^{\prime \prime}   +\rm{h.c.} \big],~~~~ \label{vl_lgr}
\end{eqnarray}
is the Yukawa interaction term for vectorlike leptons and their interactions with ordinary leptons, and
\begin{eqnarray}
V(\Phi,\Delta)&=&m^2\Phi^\dagger\Phi+M^2\rm{Tr}(\Delta^\dagger\Delta)+\left[\mu \Phi^Ti\tau_2\Delta^\dagger \Phi+\rm{h.c.}\right]+\lambda_1(\Phi^\dagger\Phi)^2 \nonumber\\
&+&\lambda_2\left[\rm{Tr}(\Delta^\dagger\Delta)\right]^2 +\lambda_3\rm{Tr}[(\Delta^\dagger\Delta)^2]
+\lambda_4(\Phi^\dagger\Phi)\rm{Tr}(\Delta^\dagger\Delta)+\lambda_5\Phi^\dagger\Delta\Delta^\dagger\Phi,~~~~ \label{eq:pot_htm}
\end{eqnarray}
is the scalar potential for the SM doublet $\Phi$ ($\langle \Phi \rangle =\frac{v_\Phi}{\sqrt{2}}$) and triplet $\Delta$ Higgs fields.  The triplet and doublet Higgs VEVs are related through $v^2=v_\Phi^2+2 v_\Delta^2\simeq (246 ~{\rm GeV})^2$. The scalar potential in Eq. (\ref{eq:pot_htm}) induces mixing among the physical states for the singly charged, the CP-odd, and the CP-even neutral scalar sectors, which are always small ${\cal O}(v_\Delta/v_\Phi)$ for the first two sectors, but not necessarily so for the latter one, 
\begin{eqnarray}
\left(
\begin{array}{c}
\varphi\\
\delta
\end{array}\right)&=&
\left(
\begin{array}{cc}
\cos \alpha & -\sin\alpha \\
\sin\alpha   & \cos\alpha
\end{array}
\right)
\left(
\begin{array}{c}
h\\
H
\end{array}\right),
\end{eqnarray}
where the mixing angle is given in terms of the parameters in $V(\Phi,\Delta)$ as
\begin{eqnarray}
\tan2\alpha &=&\frac{v_\Delta}{v_\Phi}\, \frac{2v_\Phi^2(\lambda_4+\lambda_5)-4({v_\Phi^2\mu}/{\sqrt{2}v_\Delta})^2}{2v_\Phi^2\lambda_1-({v_\Phi^2\mu}/{\sqrt{2}v_\Delta})^2-2 v_\Delta^2(\lambda_2+\lambda_3)}.~~~~~~~~ \label{tan2a}
\end{eqnarray}
In our previous work \cite{Arbabifar:2012bd}, we showed that the Higgs masses and coupling strengths are consistent with choosing $h$ to be the SM-like state at $125$ GeV, while the state $H$ a lighter state, perhaps the state observed at LEP \cite{Barate:2003sz}\footnote{This scenario was imposed by the requirement of an enhanced diphoton signal for the Higgs of mass 125 GeV, so it can be relaxed here.}. The masses of the neutral $h$ and $H$ are given by:
\begin{eqnarray}
&&m_h^2= 2v_\Phi^2\lambda_1\cos^2\alpha+\left [({v_\Phi^2\mu}/{\sqrt{2}v_\Delta})^2+2v_\Delta^2(\lambda_2+\lambda_3)\right ] \sin^2\alpha+\left [\frac{v_\Phi^3 \mu^2}{v_\Delta}-v_\Phi v_\Delta(\lambda_4+\lambda_5)\right] \sin2\alpha,\nonumber\\
&&\\
&&m_H^2=2v_\Phi^2\lambda_1\sin^2\alpha+ \left [({v_\Phi^2\mu}/{\sqrt{2}v_\Delta})^2+2v_\Delta^2(\lambda_2+\lambda_3)\right] \cos^2\alpha- \left [\frac{v_\Phi^3 \mu^2}{v_\Delta}-v_\Phi v_\Delta(\lambda_4+\lambda_5)\right ] \sin2\alpha.\nonumber\\
\end{eqnarray}
The expressions relating the $\lambda_1$-$\lambda_5$ parameters to the Higgs masses can be found in \cite{Arbabifar:2012bd}. In particular, the doubly charged Higgs boson mass is:
\begin{equation}
m_{H^{++}}^2=\frac{v_\Phi^2\mu}{\sqrt{2}v_\Delta}-v_\Delta^2\lambda_3-\frac{\lambda_5}{2}v_\Phi^2 \simeq \left (\frac{ \mu}{\sqrt{2}v_\Delta}- \frac{\lambda_5}{2} \right)v^2_{\Phi}, \label{mhpp}
\end{equation}
where we used $v_{\Delta} \ll v_{\Phi}$. As we choose $v_\Delta=1$ GeV for consistency with the value of the $\rho$ parameter, the doubly charged mass is approximately $\displaystyle m_{H^{++}} \simeq \left( \mu-\lambda_5/\sqrt{2}\right)^{1/2} (207$ GeV). The coupling $\lambda_5$ is expected to be $\le 1$ and for light doubly charged masses, the $\mu$ parameter is small\footnote{Specifically, for our parameter space $\mu=0.2$ GeV and $\lambda_5<0$.}, $\mu \sim v_\Delta \sim {\cal O}($GeV).

New symmetries can be introduced to restrict the interactions of the vector leptons. For instance,
we can impose {\it (i)} a symmetry under which all the new $SU(2)$ singlet fields are odd, while the new $SU(2)$ doublets are even, which forces all  Yukawa couplings involving new leptons to vanish,  $h_E^\prime =h_E^{\prime \prime}=h_\nu^\prime=h_\nu^{\prime \prime}=h_{ij}^{\prime}=h_{ij}^{\prime \prime}=0$, and the vector lepton masses arise only from explicit terms in the Lagrangian \cite{Joglekar:2012vc};  and/or {\it (ii)} impose a new parity symmetry which disallows mixing between the ordinary leptons and the new lepton fields, under which all the mirror fields are odd, while the others are even   \cite{Ishiwata:2013gma},  such that  $  \lambda_E^i= \lambda_L^i =\lambda_{ij}^\prime=\lambda_{ij}^\prime=\lambda_{ij}^{\prime \prime}=0$. The latter are important for light vectorlike leptons, as this scenario would satisfy restrictions from lepton-flavor violating decays, which otherwise would either force the new leptons to be very heavy, $\sim 10- 100$ TeV, or reduce the branching ratio for the Higgs into dileptons  to 30-40\% of the SM prediction. In addition, if all vectorlike leptons are odd under this symmetry, the lightest particle can become stable and act as all, or part of, the dark matter in the universe. Thus the assumption {\it (ii)}  has all the attractive features we like for this analysis, and we adopt it here,  while allowing  $h_E^\prime , h_E^{\prime \prime}, h_\nu^\prime, h_\nu^{\prime \prime}, h_{ij}^{\prime}, h_{ij}^{\prime \prime} \ne 0$. 

As we concentrate on the possibility that the lightest neutral component of the new vectorlike leptons is a dark matter candidate, we are primarily interested in light states. The  $2 \times 2$ mass matrix ${\cal M}_E$ for the charged sector is defined as \cite{Joglekar:2012vc,Bahrami:2014ska}
 \begin{eqnarray}
\left ( {\bar e}_L^\prime~~{\bar e}_L^{\prime \prime}\right ) \left ({\cal M}_E \right ) \left( \begin{array}{c} e_R^\prime\\e_R^{\prime \prime} \end{array} \right)\,, \quad {\rm with} \quad
{\cal M}_E=\left ( \begin{array}{cc}
m_E^\prime&M_L\\
M_E&m_E^{\prime\prime}\end{array} \right),  
\end{eqnarray}
with $m_E^\prime=h_E^\prime v_\Phi/\sqrt{2}$ and $m_E^{\prime \prime}=h_E^{\prime \prime} v_\Phi/\sqrt{2}$, from the Lagrangian Eq. (\ref{vl_lgr}). The mass matrix can be diagonalized by two unitary matrices $U^L$ and $U^R$ as follows:
\begin{equation}
{U^L}^\dagger {\cal M}_E U^R=\left ( \begin{array}{cc}
M_{E_1}&0\\
0&M_{E_2} \end{array} \right).
\end{equation}
The mass eigenvalues are (by convention the order is $M_{E_1}>M_{E_2}$)
\begin{equation}
M^2_{E_1, E_2}= \frac12 \left [ \left (M_L^2+m_E^{\prime\,2}+M_E^2+m_E^{\prime \prime\,2}\right ) \pm \sqrt{\left (M_L^2+m_E^{\prime\,2}-M_E^2-m_E^{\prime \prime\,2}\right )^2+4(m_E^{\prime \prime}M_L+m_E^{\prime}M_E)^2} \right ], 
\label{eq:eigenvalues}
\end{equation}
while in the neutral sector the mass matrix is:
 \begin{eqnarray}
\frac12 \left (\bar {\nu_L^\prime}~~\bar {\nu_R^{\prime \,c}}~~\bar {\nu_R^{\prime \prime\,c}}~~\bar {\nu_L^{\prime \prime}}\right ) \left ({\cal M}_\nu \right ) \left( \begin{array}{c}  \nu_L^{\prime \,c} \\ \nu_R^{\prime }\\ \nu_R^{\prime \prime} \\ \nu_L^{\prime \prime \,c} \end{array}  \right )\, , \quad {\rm with} \quad
{\cal M}_\nu=\left ( \begin{array}{cccc}
0&m_\nu^\prime& M_L &0\\
m_\nu^\prime&M_\nu^{\prime}&0&M_\nu \\
M_L&0&0&m_\nu^{\prime \prime} \\
0&M_\nu&m_\nu^{\prime \prime}&M_{\nu}^{\prime \prime}
\end{array} \right),  
\end{eqnarray}
with $m_\nu^\prime=h_\nu^\prime v_\Phi/\sqrt{2}$ and $m_\nu^{\prime \prime}=h_\nu^{\prime \prime} v_\Phi/\sqrt{2}$. This mass matrix can be diagonalized  by a unitary matrix $V$:
\begin{equation}
{V}^\dagger {\cal M}_\nu V=\left ( \begin{array}{cccc}
M_{\nu_1}&0&0&0\\
0&M_{\nu_2}&0&0 \\
0&0&M_{\nu_3}&0\\
0&0&0&M_{\nu_4}   \end{array} \right). 
\end{equation}
In the limit where the explicit mass terms $M_L$, $M_E$ and $M_\nu$ in the interaction Lagrangian vanish, after electroweak symmetry breaking  there are
two charged leptons with masses $m_E^\prime$ and $m_E^{\prime \prime}$, and four neutrinos with masses:
\begin{eqnarray}
M_{\nu_{1,2}}&=& \sqrt{\frac{M_\nu^{\prime \, 2}}{4}+m_\nu^{\prime \,2}} \pm \frac{M_\nu^\prime}{2}\, \\
M_{\nu_{3,4}}&=& \sqrt{\frac{M_\nu^{\prime \prime \, 2}}{4}+m_\nu^{\prime \prime \,2}} \pm \frac{M_\nu^\prime}{2}.
\end{eqnarray}
The lightest of these eigenvalues will be the dark matter candidate, and as it is odd under the additional parity symmetry {\it (ii)}, it is stable. For vanishing $h_\nu^\prime,~h_\nu^{\prime \prime}$ Yukawa couplings, the two singlet vectorlike neutrinos have vanishing couplings with the $Z$ boson. Lifting the Yukawa couplings slightly from 0 allows mixing between the singlet neutrinos and the neutral components of the doublet vectorlike leptons, inducing a (small) coupling to the $Z$ boson. For simplicity,  we adopt the scenario in  \cite{Joglekar:2012vc} where $h_\nu^\prime \ne 0$, but $h_{\nu}^{\prime \prime}=0$, as well as setting the explicit neutrino mass in the Lagrangian $M_\nu=0$. This scenario is sufficient to provide a single DM candidate and a single Yukawa coupling, and transparent enough to yield consequences. It corresponds to one neutrino state which does not mix and is sterile  (the mirror $SU(2)_L$ doublet $\nu_L^{\prime \prime}$), while the remaining neutral sector consists of three neutrinos which mix, with mixing matrix in the $(\nu_L^{\prime \,c}, \nu_R^\prime, \nu_R^{\prime \prime})$ basis given by 
\begin{eqnarray}
{\cal M}_{3 \nu}=\left ( \begin{array}{ccc}
0&m_\nu^\prime& M_L \\
m_\nu^\prime&M_\nu^{\prime}&0 \\
M_L&0&0 
\end{array} \right).  
\end{eqnarray}
The Yukawa coupling $h_\nu^\prime$ must remain small  to insure smallness of couplings to the $Z$ boson. In the limit $h_\nu^\prime=0$, the matrix has two degenerate eigenvalues of mass $M_L$, predominantly $SU(2)_L$ doublets, and one state with mass $M_\nu^{\prime }$ and predominantly singlet. For $h_\nu^\prime \ne 0$, these three states mix, generating a small mixing coupling to the $Z$ boson.  The lightest neutrino state $M_{\nu_1}$ emerges as being dominantly  $\nu_R^\prime$ and is the dark matter candidate. For the charged lepton sector, we take $M_L= 205$ GeV and $M_E= 300$ GeV and $h_E^{\prime}=h_E^{\prime \prime}=0.8$ \cite{Joglekar:2012vc}. In this case the lightest charged lepton will be $M_{E_2}\sim 108$ GeV, close to the LEP limit, $M_{E}>102.6$ GeV \cite{pdg}, which imposes an upper limit on the mass of the dark matter candidate $M_{DM}\equiv M_{\nu_1} < M_{E_2}$. 

Next, we analyze the effects of the new states on electroweak precision parameters in the HTM and consequently, the restrictions imposed on its parameter space.

%
%


\section{Vectorlike lepton contributions to the $S$ and $T$ parameters}
\label{sec:STU}
Adding new particles to the model spectrum affects quantum corrections on the propagators of $W$ and $Z$ bosons. The corrections are parametrized by two oblique parameters,  $S$ and $T$ \footnote{We set $U=0$.}, which encapsulate the model restrictions coming from electroweak precision data. For a Higgs state with mass $m_h=125$ GeV, the allowed ranges are  \cite{Kanemura:2012rs}
\begin{eqnarray}
&&\Delta S=S-S_{\rm SM}= 0.05 \pm 0.09,\nonumber\\
&&\Delta T=T-T_{\rm SM}= 0.08 \pm 0.07\, ,
\end{eqnarray}
with errors correlated by a factor of 0.88.
The explicit expressions for the $S$, $T$ and $U$ parameters for the HTM are given in \cite{Bahrami:2014ska}. The addition of vectorlike leptons modifies these by the following contributions. For the $S$ parameter \cite{Joglekar:2012vc}:
\begin{eqnarray} 
S &=& \frac{1}{\pi} \left \{ \sum_{j,k=1}^2  (|U_{1 j}^L|^2|U_{1 k}^L|^2+ |U_{2 j}^R|^2|U_{2 k}^R|^2)\, b_2(M_{E_j},M_{E_k},0)
+ \sum_{j,k=1}^2 { \rm Re}(U_{1 j}^L U_{1 k}^{L\,\star} U_{2 j}^{R\,\star}U_{2 k}^R)f_3(M_{E_j},M_{E_k}) \right. \nonumber \\
&+& \left. 
 \sum_{j,k=1}^3  (|V_{1 j}|^2|V_{1 k}|^2+ |V_{3 j}|^2|V_{3 k}|^2) \, b_2(M_{\nu_j},M_{\nu_k},0) 
+  \sum_{j,k=1}^3 { \rm Re}(V_{1 j} V_{1 k}^{\star} V_{3 j}V_{3 k}^{\star})f_3(M_{\nu_j},M_{\nu_k})  \right. \nonumber \\
 &-& \left. 2\sum_{j=1}^2  (|U_{1 j}^L|^2+ |U_{2j}^R|^2 )\, b_2(M_{E_j},M_{E_j},0)  + \frac{1}{3} \right \} \, , 
\end{eqnarray}
while the oblique correction parameter $T$ for vectorlike leptons is \cite{Joglekar:2012vc}:
\begin{eqnarray} 
T &=& \frac{1}{4 \pi s^2_W c^2_W M^2_Z} \left \{ -2\sum_{j=1}^{2} \sum_{k=1}^{3} (|U_{1 j}^L|^2|V_{1 k}|^2+ |U_{2 j}^R|^2|V_{3 k}|^2) \, b_3(M_{\nu_k},M_{E_j},0) \right. \nonumber \\
&+& \left.  2\sum_{j=1}^{2} \sum_{k=1}^{3} { \rm Re}(U_{1 j}^L U_{2 j}^{R\,\star} V_{1 k} V_{3 k}^\star )M_{E_j} M_{\nu_k} b_0(M_{E_j},M_{\nu_k},0)  
\right. \nonumber \\
&+& \left. \sum_{j,k=1}^3  (|V_{1 j}|^2|V_{1 k}|^2+ |V_{3 j}|^2|V_{3 k}|^2)\,  b_3(M_{\nu_j} M_{\nu_k},0)  \right. \nonumber \\
&-& \left.  \sum_{j,k=1}^3 { \rm Re}(V_{1 j} V_{1 k}^{\star} V_{3 j}V_{3 k}^{\star}) M_{\nu_j} M_{\nu_k} b_0(M_{\nu_j},M_{\nu_k},0) \right. \nonumber \\
&+& \left.  (|U_{1 1}^L|^4 +|U_{2 1}^R|^4) M^2_{E_1}  b_1(M_{E_1},M_{E_1},0)+ (|U_{1 2}^L|^4 + |U_{2 2}^R|^4) M^2_{E_2}  b_1(M_{E_2},M_{E_2},0) \right. \nonumber \\
&+& \left.  (2|U_{1 1}^L|^2|U_{2 1}^L|^2+2|U_{1 2}^R|^2|U_{2 2}^R|^2)\,  b_3(M_{E_1},M_{E_2},0) \right. \nonumber \\
 &-& \left. \sum_{j,k=1}^2  { \rm Re}(U_{1 j}^L U_{1 k}^{L\,\star} U_{2 j}^{R\,\star}U_{2 k}^R) M_{E_j} M_{E_k} b_0(M_{E_j},M_{E_k},0)  \right \} \, ,
\end{eqnarray}
where  the Passarino-Veltmann functions are: 
\begin{eqnarray}
b_{0}(M_1, M_2, q^2) &=& {\int_0}^1 \log(\frac{\Delta}{\Lambda^2})dx ,\\
b_{1}(M_1, M_2, q^2) &=& {\int_0}^1 x\log(\frac{\Delta}{\Lambda^2})dx, \\
b_{2}(M_1, M_2, q^2) &=& {\int_0}^1 x (1-x)\log(\frac{\Delta}{\Lambda^2})dx=b_{2}(M_2, M_1,q^2),\\
b_{3}(M_1, M_2, 0) &=& \frac{M^2_2 b_1(M_1,M_2,0)+M^2_1 b_1(M_2,M_1,0)}{2},\\
f_{3}(M_1, M_2) &=& M_1 M_2\frac{M^4_2-M^4_1+2 M^2_1 M^2_2 \log(\frac{M^2_1}{M^2_2})}{2(M^2_1-M^2_2)^3}.
\end{eqnarray}
We defined  $\Delta = M_2^2x+M_1^2 (1-x)-x(1-x)q^2$ and in the above $\Lambda^2$ is an arbitrary regularization scale that will not affect physical observables. The function $f_{3}(M_1, M_2)=-1/6$ remains well defined in the limit $M_2\rightarrow M_1$. 
 As in the HTM without vectorlike leptons, the $S$ parameter does not impose any restrictions on the parameter space of the model, while the $T$ parameter is very restrictive. The reason is that $T$ depends quadratically on mass differences, while $S$ only logarithmically.

We  proceed to analyze restrictions on the relevant masses and couplings in the model coming from the $T$ parameter. In Fig. \ref{fig:Tmhppsin} we show the effects on the $T$ parameter as a contour in a $m_{H^{\pm \pm}}-\sin \alpha$ plane (with $\sin \alpha$  the mixing angle in the neutral Higgs sector), for two values of dark matter masses, $M_{DM}=30$ GeV and $M_{DM}=50$ GeV.  The allowed values for this parameter, $-0.2 < \Delta T< 0.4$, are given in the code bars (colored contours in the figure). The maximum doubly charged mass values allowed are $m_{H^{\pm\pm}} \sim (280-290)$ GeV for dark matter masses in  24-30 GeV and 70-90 GeV regions, and approximately 250-270 GeV for dark matter masses in 30-70 and 90-103 GeV range. We have selected the particular values for $M_{DM}=30$ GeV and $h_\nu^\prime=0.65$ (left panel) and  $M_{DM}=50$ GeV and $h_\nu^\prime=0.28$ (right panel) to belong to the parameter space where  the relic density is within experimental bounds, as explained in detail in Sec.  \ref{sec:RD}.  As the figure indicates, the $T$ parameter depends only slightly on $\sin \alpha$, but is extremely sensitive to the mass of the doubly charged Higgs boson.  

%
%
\begin{figure}[htbp]
\vskip -0.3in
\begin{center}$
    \begin{array}{cc}
\hspace*{-0.5cm}
    \includegraphics[width=3.4in,height=3in]{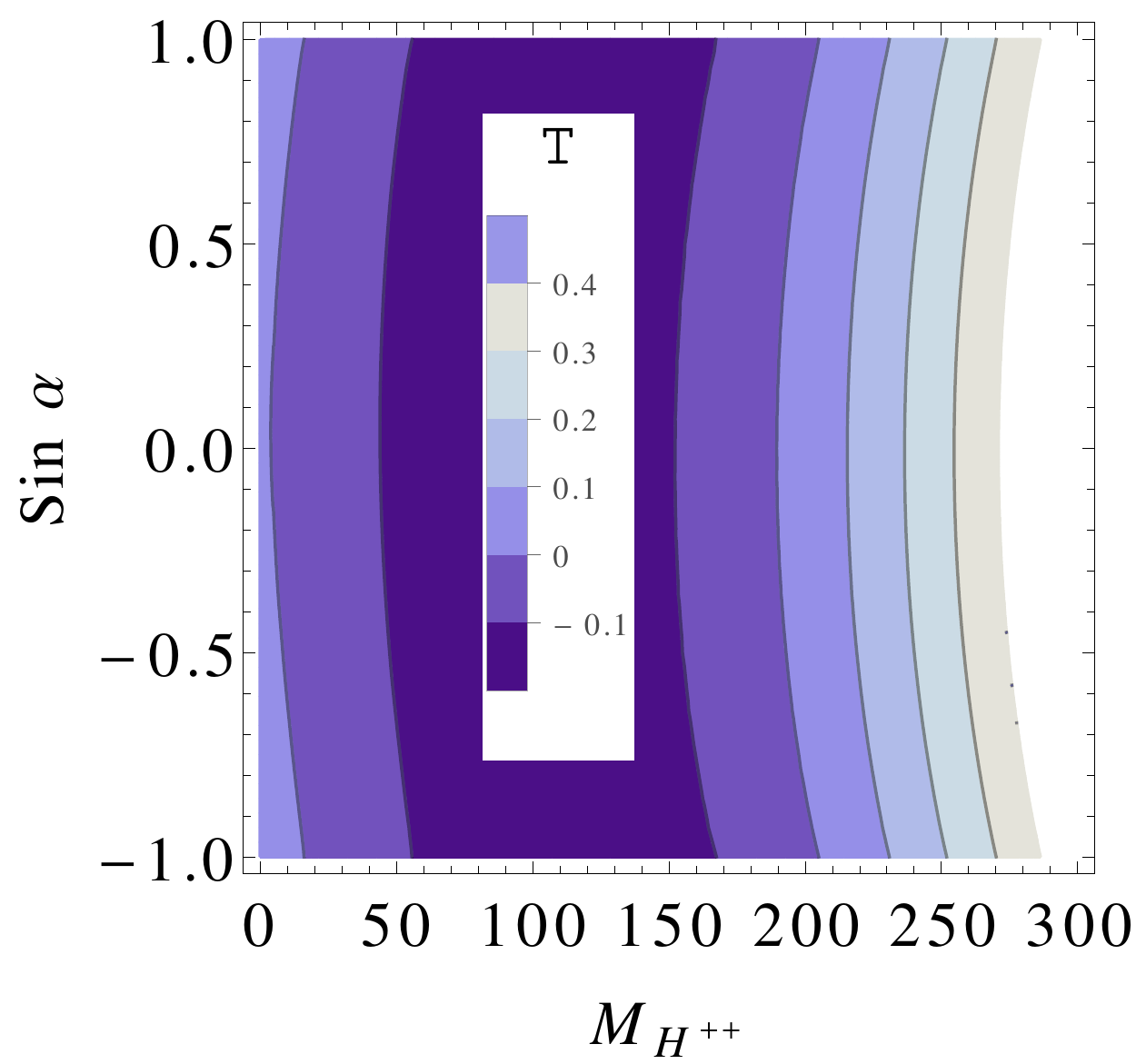}
&\hspace*{-0.2cm}
    \includegraphics[width=3.4in,height=3in]{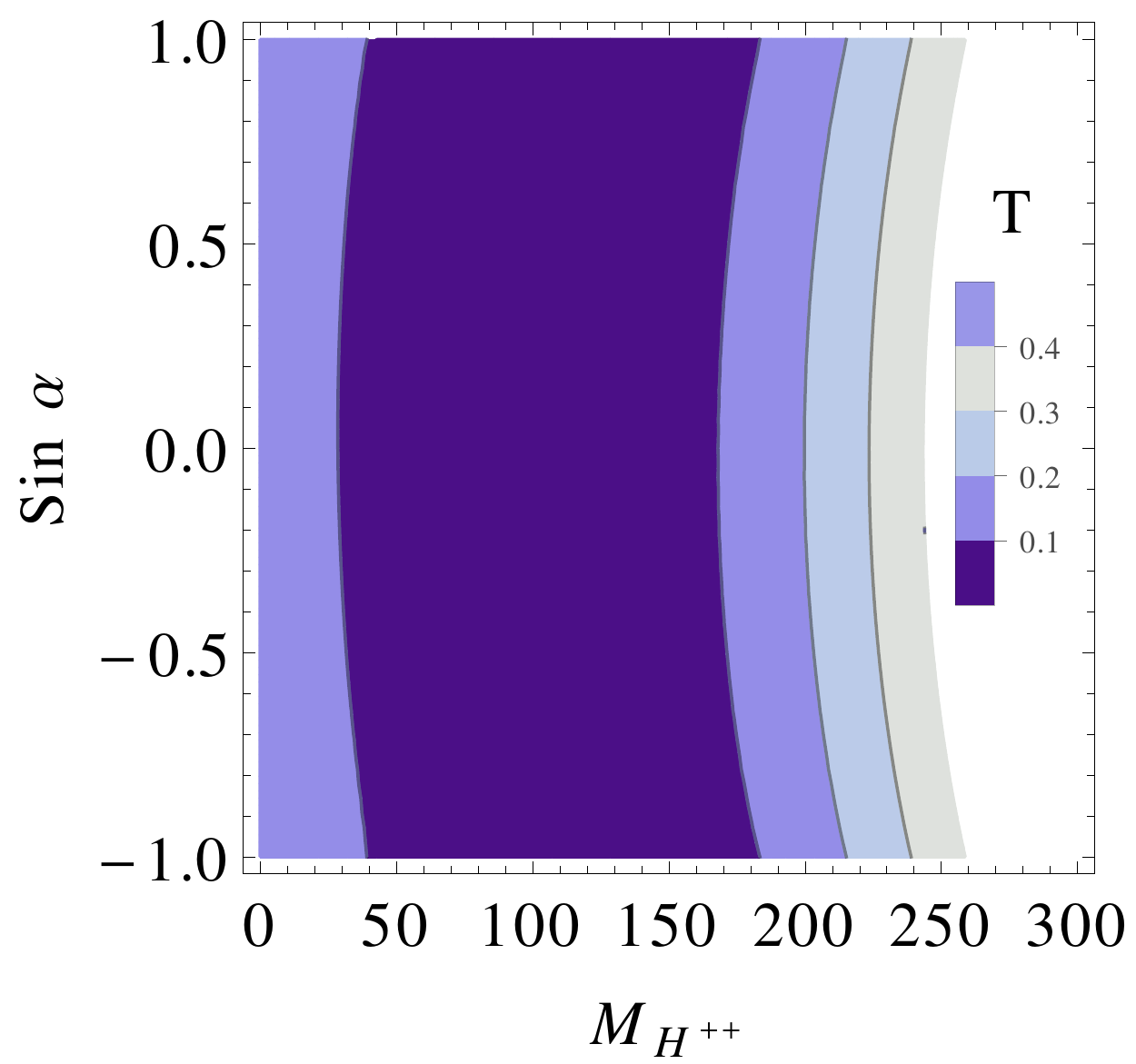}
        \end{array}$
\end{center}
\vskip -0.2in
     \caption{(color online). \sl\small Contour graphs showing the contribution to the $T$  parameter in the HTM (as given in the code bars) with vectorlike leptons, as a function of the doubly charged Higgs mass $m_{H^{\pm\pm}}$ and the mixing angle $\sin \alpha$, for fixed values of the neutrino Yukawa coupling $h_\nu^\prime$. We take (left panel)  $M_{DM}=30$ GeV, $h_\nu^\prime=0.65$, (right panel) $M_{DM}=50$ GeV, $h_\nu^\prime=0.28$. The allowed range of T parameter is $-0.2 < \Delta T< 0.4$. The white region represents the parameter region ruled out by the constraints. }
\label{fig:Tmhppsin}
\end{figure}
\begin{figure}[htbp]
\begin{center}$
    \begin{array}{ccc}
\hspace*{-1.7cm}
    \includegraphics[width=2.5in,height=2.8in]{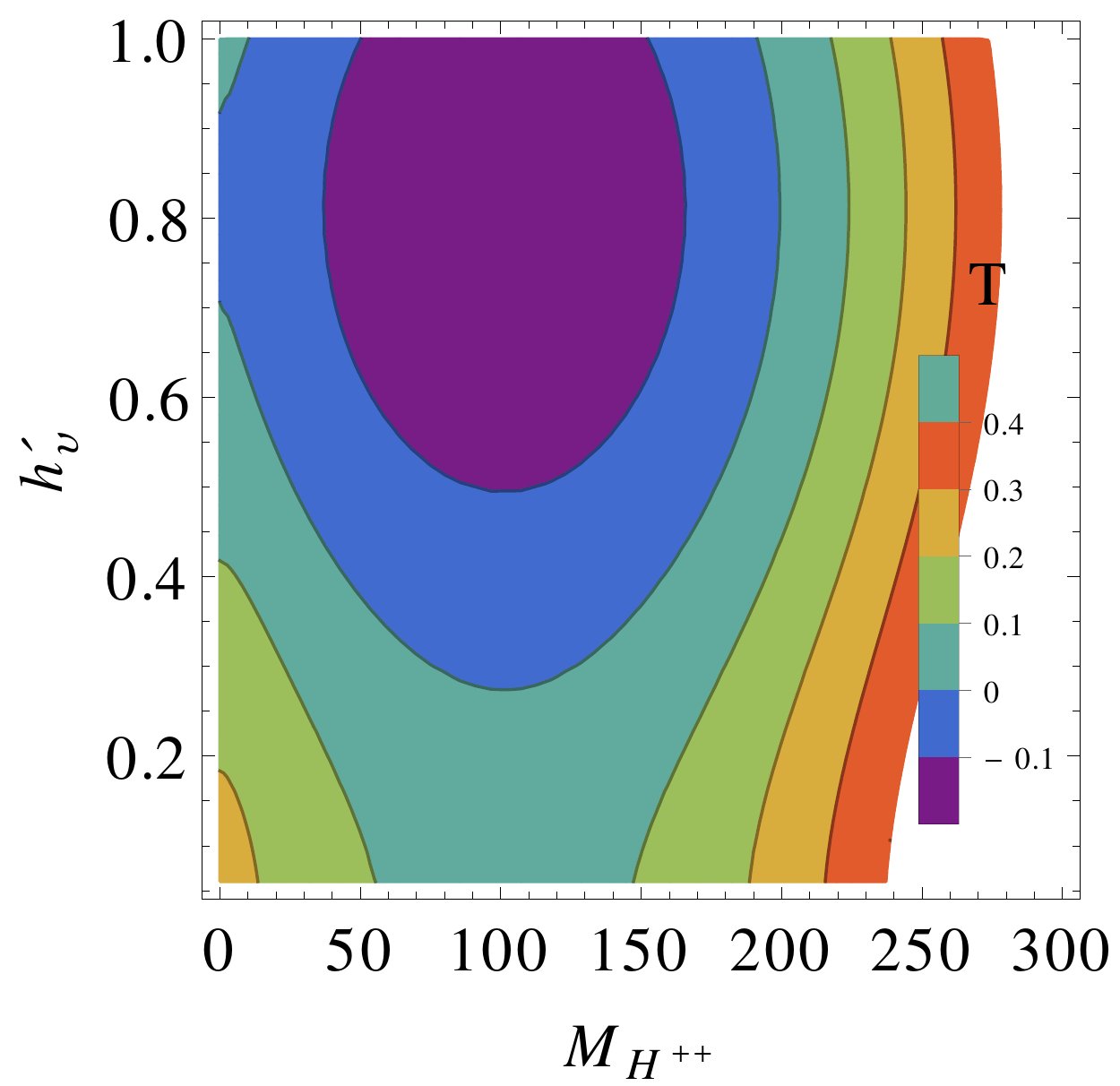}
&\hspace{-0.2cm}
\includegraphics[width=2.5in,height=2.8in]{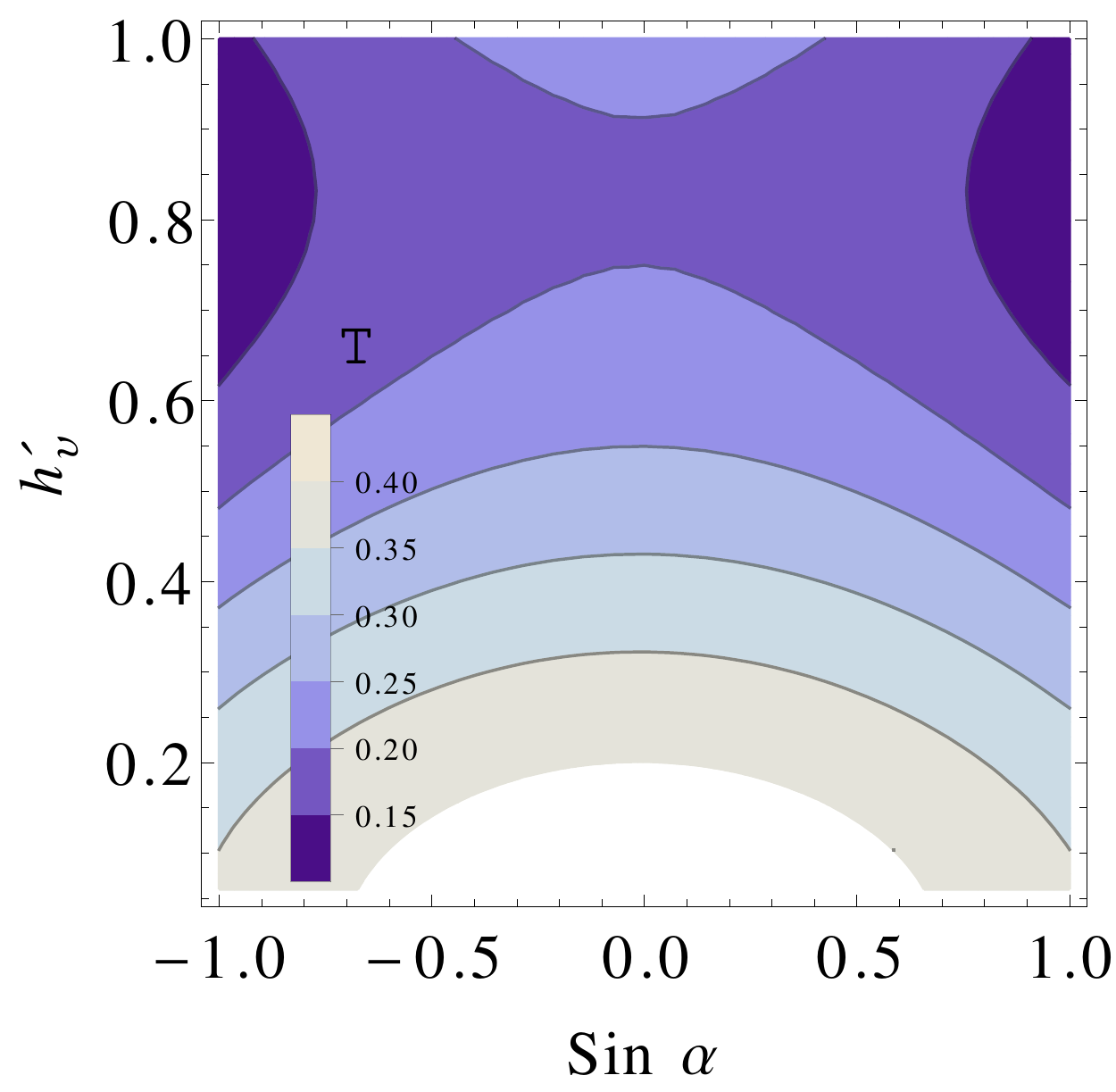}                                                                             
& \hspace{-0.2cm}
\includegraphics[width=2.5in,height=2.8in]{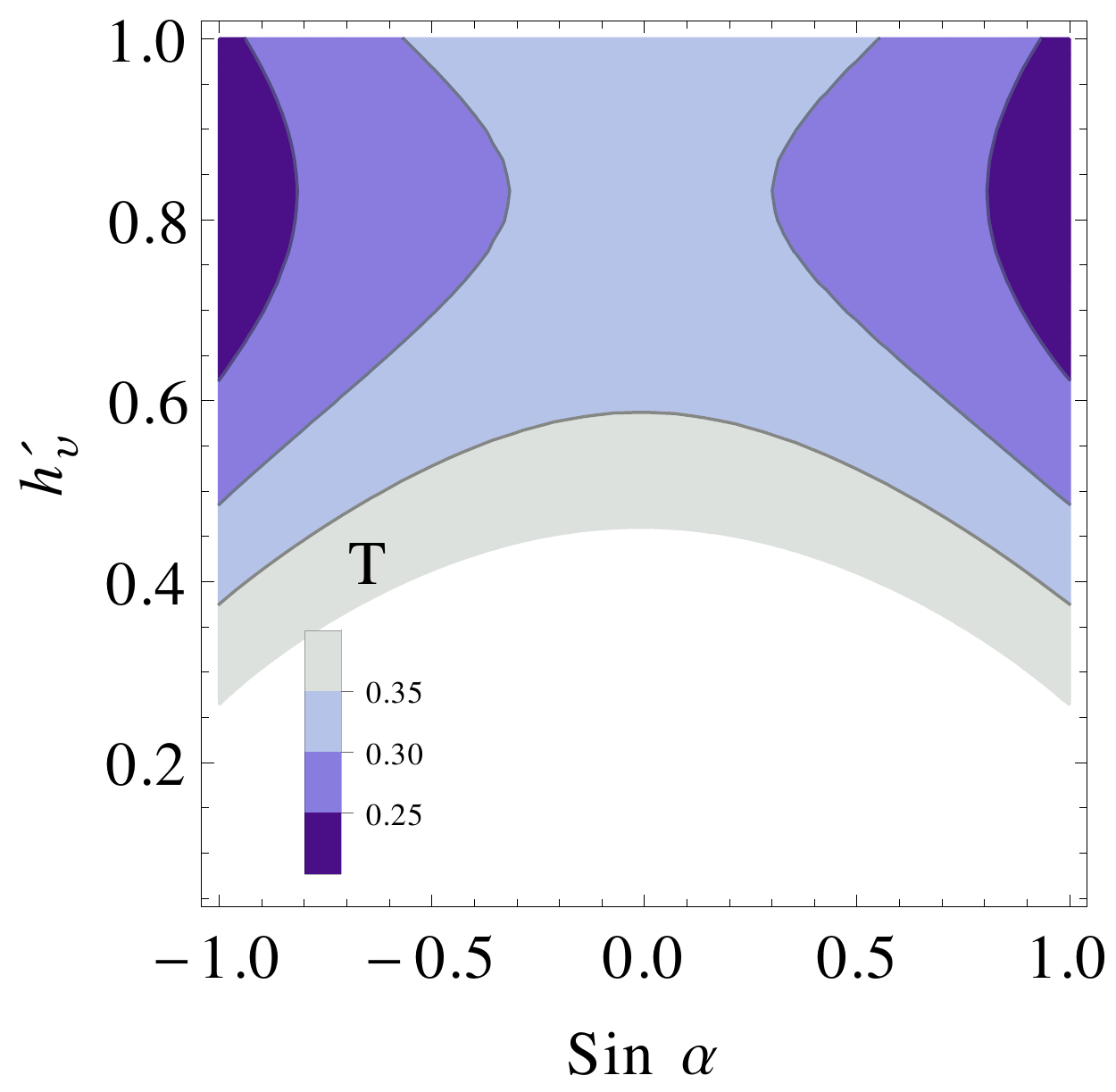}
        \end{array}$
\end{center}
\vskip -0.2in
     \caption{(color online). \sl\small Contour graphs showing the contribution to the $T$ parameter in the HTM with vectorlike leptons (values as given in the code bars, within the allowed range  $-0.2 < \Delta T< 0.4$.) as a function of the parameters of the model.  In the left plot, we show the combined dependence on $h_\nu^\prime$ and  $m_{H^{\pm\pm}}$ (for $M_{DM}=30$ GeV and $\sin \alpha=0.5$); in the middle (right)  panel,  the dependence on $h_\nu^\prime$ and $\sin \alpha$ for $M_{DM}=50$ GeV  and $m_{H^{\pm\pm}}=240$ GeV ($m_{H^{\pm\pm}}=260$ GeV).  The white region represents the parameter region ruled out by the constraints.}
\label{fig:TmDMyn}
\end{figure}
In Fig. \ref{fig:TmDMyn} we show the variation of the $T$ parameter as a contour in an $m_{H^{\pm\pm}}-h_\nu^\prime$ plane (left panel) and $h_\nu^\prime-\sin \alpha$ (middle and right panels). For the left panel, we chose an illustrative example with $M_{DM}=30$ GeV and $\sin \alpha=0.5$; the $T$ parameter does not depend sensitively on varying these, but again it is very sensitive to the mass of the doubly-charged Higgs boson, as shown in the middle and right side panels, where increasing the value of $m_{H^{\pm\pm}}$ from 240 to 260 GeV places significant restrictions on the $T$ parameter. Increasing the mass of the doubly charged Higgs boson and decreasing $\sin \alpha$ (the mixing angle) impose restrictions on $h_\nu^\prime$ (the vectorlike neutrino Yukawa coupling) from the $T$ parameter.    Note that the  $T$ parameter is not sensitive to the mass of the dark matter candidate and it affects it only indirectly, through the restrictions on the Yukawa couplings.

\section{Invisible decay width of the Higgs boson}
\label{sec:IDW}
%
The existence of the vectorlike neutrino $\nu_1$ as dark matter candidate will have an effect on the branching ratio of the Higgs boson, if $m_h\ge 2M_{\nu_1}$. Given that $\nu_1$ is stable, the decays $h \to \nu_1 \nu_1,\, h \to \nu_1 {\bar \nu}_1$ will contribute to the invisible Higgs branching ratio, which is constrained by combined  CMS and ATLAS measurements to be $BR_{\rm inv}<58\%$ for a SM Higgs with a mass of 125 GeV \cite{Lange:2014fxa}, and more stringently by global fits to be $BR_{\rm inv}$ of 29\% with 95\% C.L.  \cite{Belanger:2013xza}.

 In the Higgs Triplet Model, the tree-level decay width of the Higgs boson into vectorlike neutrinos is \cite{Arbabifar:2012bd,Heikinheimo:2012yd}
\begin{eqnarray}
\label{eq:invisible}
[\Gamma(h \rightarrow \nu_1 \bar{\nu_1})]_{HTM}
& = & \frac{G_F m_h(M_{\nu_1}C^h_{\nu_1\bar{\nu_1}})^2}{2\pi \sqrt{2}} \bigg(1-(\frac{2M_{\nu_1}}{m_{h}})^2\bigg)^{\frac{3}{2}} \cos^2\alpha\, , 
\end{eqnarray}
where 
\begin{equation}
C^h_{\nu_1 \bar{\nu_1}}=\sqrt{2} h_\nu^\prime Re(V_{1 1}V_{2 1})
\label{eq:coupling}
\end{equation}
is the Higgs coupling to the lightest vectorlike neutrino ($\nu_1$). As well, the component from the  neutral triplet Higgs field violates lepton number  and can decay into two neutrinos as
\begin{eqnarray}
\label{eq:fermion}
[\Gamma(h \rightarrow \nu_1 \nu_1)]_{HTM}&\equiv & \Gamma(h \to \nu_1^c \bar{\nu_1}) + \Gamma(h \to \bar {\nu_1^c} {\nu_1})\nonumber \\
& = & \frac{1}{2}|h^\prime_{\nu_1 \nu_1}|^2 \frac{m_{h}}{4 \pi}\left ( 1-2\frac{M_{\nu_1}^2}{m_{h}^2} \right ) \left ( 1-4\frac{M_{\nu_1}^2}{m_{h}^2} \right )^2\sin^2\alpha\, ,
\end{eqnarray}
where $h^\prime_{\nu_1 \nu_1}$ is the triplet coupling constant from Eq. (\ref{vl_lgr}).
The invisible branching ratio of the Higgs boson is defined as
\begin{equation}
BR_{\rm inv} =  \frac{[\Gamma(h \rightarrow \nu_1 \bar{\nu_1})]_{HTM}+[\Gamma(h \rightarrow \nu_1 \nu_1)]_{HTM}}{
[\Gamma(h \rightarrow \nu_1 \bar{\nu_1})]_{HTM}+[\Gamma(h \rightarrow \nu_1 \nu_1)]_{HTM}+
[\Gamma(h)]_{HTM}},
\label{eq:ratio}
\end{equation}
where $[\Gamma(h)]_{HTM}$ is the total Higgs decay width in the HTM without vectorlike leptons. 

%

In Fig. \ref{fig:invisible}, we show the invisible branching ratio of Higgs boson ($BR_{\rm inv}$) in the HTM with vectorlike leptons as a contour plot in an $M_{DM} - h_\nu^\prime$ plane,  for triplet Yukawa coupling $h^\prime_{\nu_1 \nu_1}=0.01$. We compare the calculation with  the upper limit on  $BR_{\rm inv}$ 
derived from global fits to ATLAS and CMS data \cite{Belanger:2013xza}\footnote{These global fits, though more restrictive, are completely consistent with our analyses and do not restrict the parameter space unnecessarily.}. As expected, the region restricted is only for $M_{DM}<m_h/2$, where
the Higgs can decay to pairs of dark matter with a sizeable width. The left panel depicts the invisible width for the mixing angle in the neutral CP-even Higgs sector, $\sin \alpha=0.1$, the  middle panel for $\sin \alpha=0.5$ and the right panel for $\sin \alpha=0.8$. The figures show that increasing the Yukawa coupling ($h_\nu^\prime$) results in an increase the invisible branching ratio of Higgs boson ($BR_{\rm inv}$) as the decay into DM is enhanced, while decreasing $\sin \alpha$ imposes more restrictions on $h_\nu^\prime$ in order to get correct $BR_{\rm inv}$, indicating that both the doublet and  triplet Higgs components play an important role in the invisible decay.

\begin{figure}[htbp]
\begin{center}$
    \begin{array}{ccc}
\hspace*{-1.0cm}
    \hspace*{-0.2cm}
    \includegraphics[width=2.3in,height=2.7in]{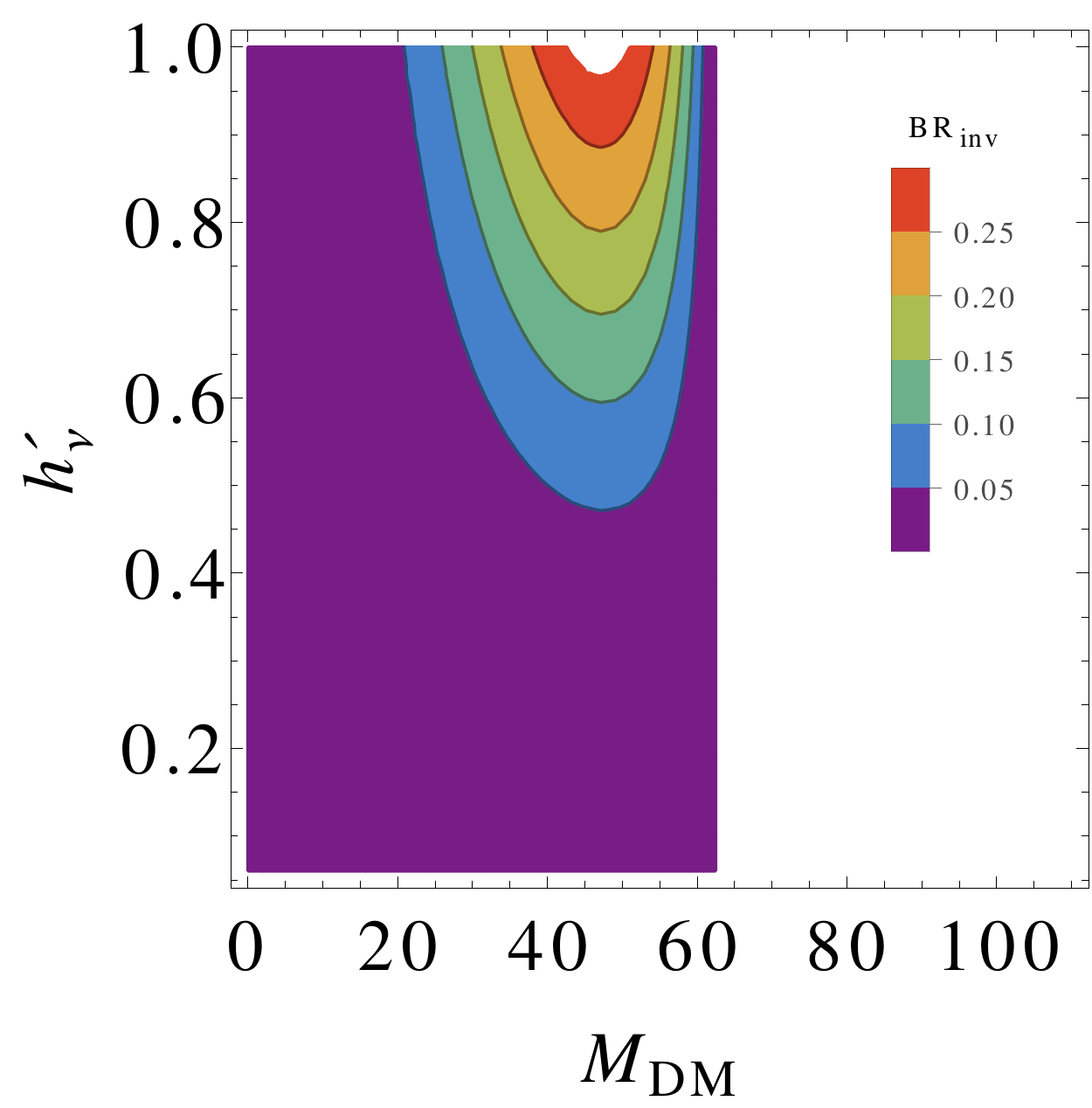}
&\hspace{-0.2cm}
\includegraphics[width=2.3in,height=2.7in]{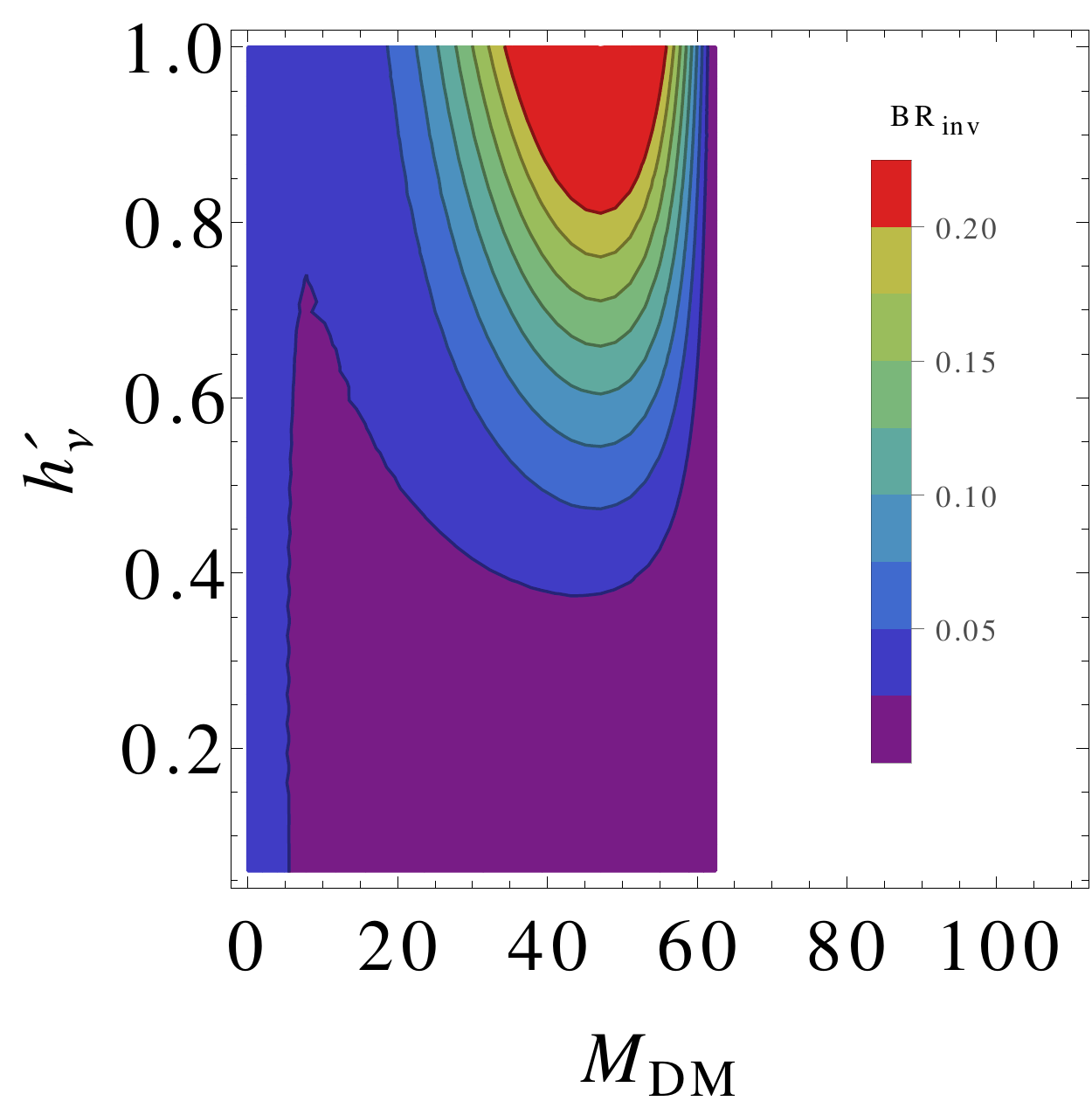}
   & \hspace{-0.2cm}
   \includegraphics[width=2.3in,height=2.7in]{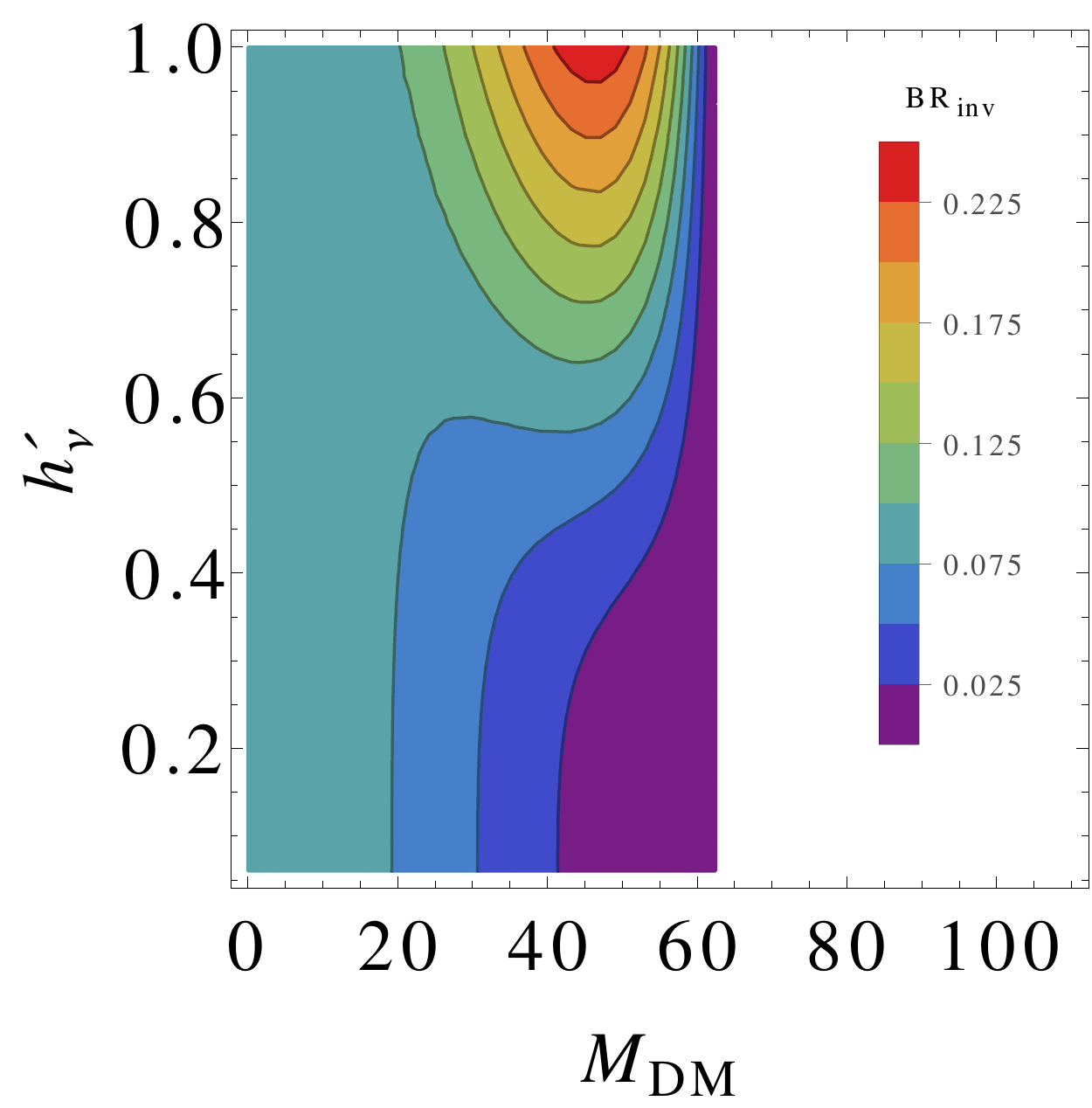}\\
        \end{array}$
\end{center}
\vskip -0.2in
     \caption{(color online). \sl\small Contour graphs showing the invisible branching ratio of Higgs boson ($BR_{\rm inv}$) in the HTM with vectorlike leptons, as functions of the dark matter mass $M_{DM}=M_{\nu_1}$ (GeV) and the neutrino Yukawa coupling $h_\nu^\prime$,  for $h^\prime_{\nu_1 \nu_1}=0.01$ . We compare to the upper limit of $BR_{\rm inv}=29$ \% from global fits to ATLAS and CMS data \cite{Belanger:2013xza} and we chose $\sin \alpha=0.1$ (left panel), $\sin \alpha=0.5$ (middle panel), $\sin \alpha=0.8$ (right panel). If the mass of DM neutrino is in the white region, it does not contribute to the Higgs invisible decay width. }
\label{fig:invisible}
\end{figure}

\section{Dark Matter Relic Density}
\label{sec:RD}
%
Global fits to a number of cosmological data (Cosmic Microwave Background, Large Scale Structure and Type Ia Supernovae) determine very precisely the amount of non-baryonic (DM) in the energy-matter of the universe at $\Omega_{DM}h^2=0.1123\pm 0.0035$ \cite{Komatsu:2010fb}, where $\Omega_{DM}$ is the energy density of the DM with respect to the critical energy density of the universe, and $h$ is the reduced Hubble parameter. Any analysis of DM must correctly replicate this value.  

To this end, we used  {\tt CalcHEP} \cite{Belyaev:2012qa} to implement the Lagrangian of the HTM with vectorlike leptons into {\tt micrOMEGAs} \cite{Belanger:2013oya}, to calculate the relic density ($\Omega_{DM} h^2$), spin-dependent cross section ($\sigma^{SD}$), spin-independent cross section ($\sigma^{SI}$), annihilation cross section ($\langle \sigma v \rangle$), and the flux of neutrino and muon predicted by the model.  For the purpose of comparing with the data, we consider the $2\sigma$ allowed range of relic density: $0.1144 \leq \Omega_{DM} h^2 \leq 0.1252$, as constrained by WMAP \cite{Komatsu:2010fb} and PLANCK \cite{Ade:2013zuv}.

In Fig. \ref{RD}, we present the allowed range of relic density of dark matter as function of the dark matter mass $M_{DM}$ (GeV) and the Yukawa coupling $h_\nu^\prime$, for two different values of the mixing angle, $\sin \alpha=0$ (left panel) and $\sin \alpha=0.8$ (right panel). Due to resonant annihilation into $Z$ bosons or Higgs boson $h$ respectively, we can see two dips at $M_{DM} \sim 45$  GeV and $M_{DM} \sim 62$ GeV. For a fixed Yukawa coupling ($h_\nu^\prime$) the cross sections becomes enhanced at the $Z$ pole and similarly at the Higgs pole, with a dominant decay into quark/antiquark.
  As the dark matter relic density is inversely proportional to the annihilation cross section, the relic density decreases in these regions. Thus, in order to produce the correct dark matter relic density, we need to decrease the value of Yukawa coupling $h_\nu^\prime$ to compensate for the effects of $Z$ and $h$ resonances, which produces the two dips at $M_{DM}=M_Z/2$ and $M_{DM}=m_h/2$. Above $M_{DM}=80$ GeV annihilation into $W^+ W^-$ pairs (and later also $Z$ bosons) becomes kinematically accessible. Finally, the relic density becomes dramatically suppressed for $M_{DM} \sim 100$  GeV due to co-annihilation with the lightest charged vectorlike lepton \cite{Joglekar:2012vc,Heikinheimo:2012yd}.  The effect of the Higgs resonance at $M_{DM} \sim 62$ GeV is slightly more pronounced for $\sin \alpha=0.8$ than for $\sin \alpha=0$  (this is the effect of increasing the triplet component contribution) and, above $M_{DM}=80$ GeV, for $\sin \alpha=0.8$, the relic abundance decrease is slightly more pronounced than in the case with $\sin \alpha=0$, but the changes are small. Overall, the graph for $\sin \alpha=0.8$ shows no marked difference from the one with $\sin \alpha=0$. The results shown are for $m_{H^{\pm\pm}}=240$ GeV. We calculated relic density for different values of the doubly-charged Higgs boson mass and found that it is insensitive to variations in this parameter. Relic density constraints restrict the dark matter mass to be heavier than 23 GeV and lighter than 103 GeV in our model, independent of any other parameters, such as Yukawa couplings or mixing angles.

\begin{figure}[htbp]
\begin{center}$
    \begin{array}{cc}
\hspace*{-0.5cm}
    \includegraphics[width=3.3in,height=3in]{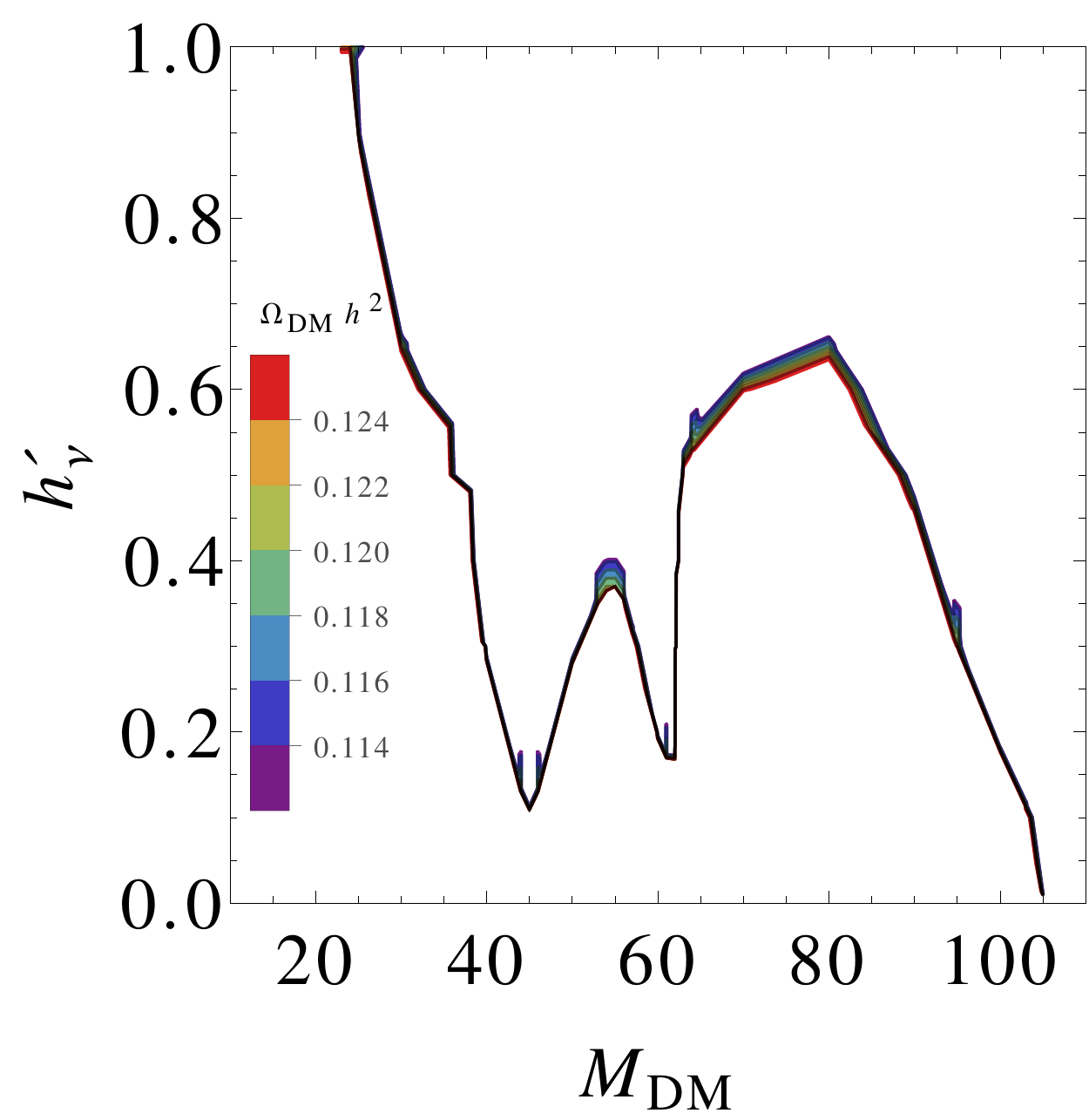}
&\hspace*{-0.2cm}
    \includegraphics[width=3.3in,height=3in]{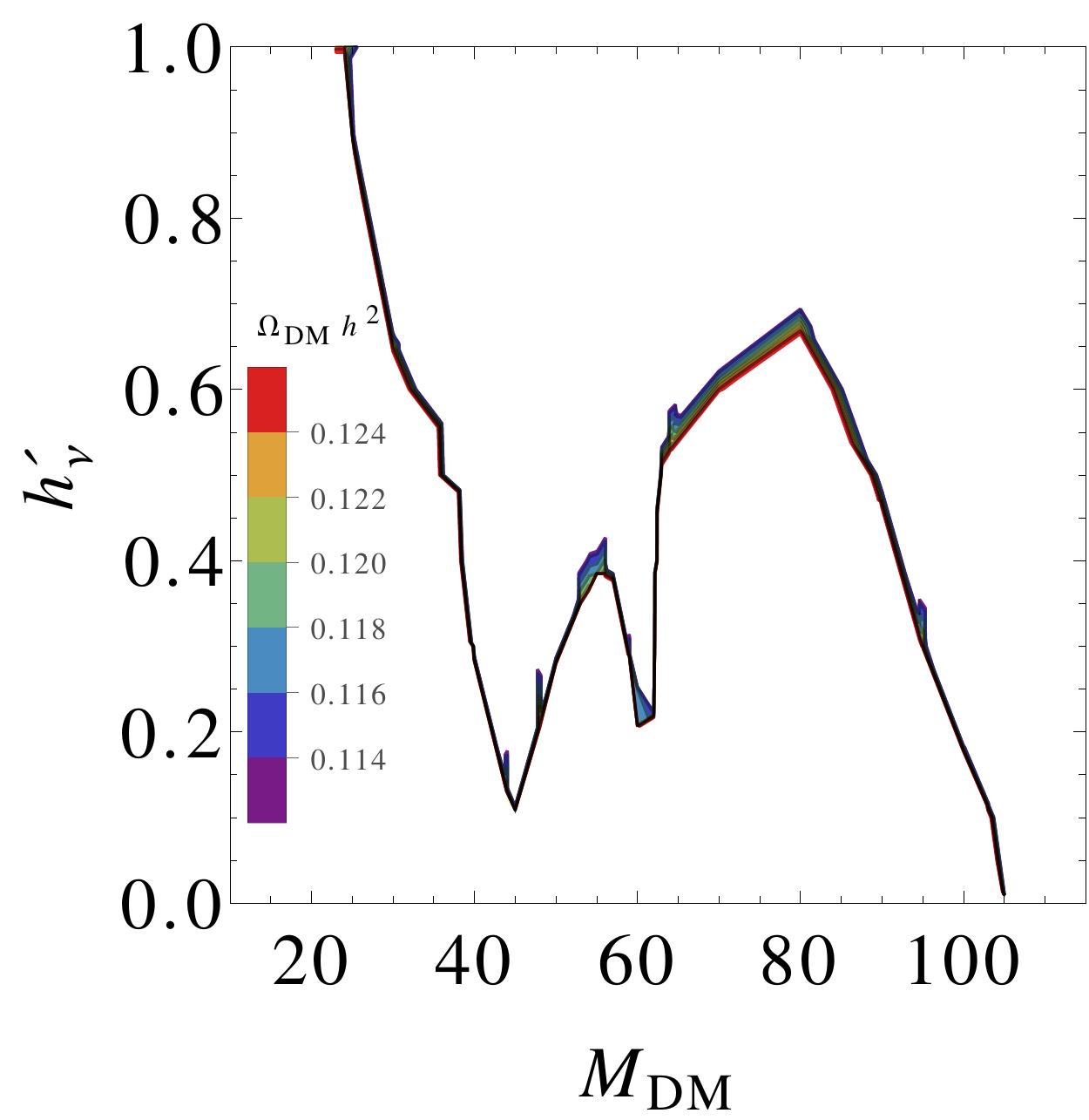}
        \end{array}$
\end{center}
\vskip -0.2in
\caption{(color online). \sl\small Contour graphs showing the correct relic density of dark matter as function of the dark matter mass $M_{DM}$ (in GeV) and the neutrino Yukawa coupling $h_\nu^\prime$ in the HTM with vectorlike leptons for $\sin \alpha=0$ (left panel) and $\sin \alpha=0.8$ (right panel). We impose the restriction $0.1144 \leq \Omega_{DM} h^2 \leq 0.1252$. The relic density is insensitive to the doubly charged Higgs boson mass, chosen here to be 240 GeV.  } 
\label{RD}
\end{figure}

\section{Direct Detection}
\label{sec:DD}
%
Dark Matter is spread over the whole universe. This provides the opportunity to detect it as it passes through and scatters off normal matter (neutrons or protons), producing detectable signals. Though direct detection is the most straightforward method of detecting DM, such events are very rare,  deposited energies very small, and thus direct detection requires very sensitive detectors with highly accurate background rejection. The expected signals depend on the nature of the DM. For vectorlike neutrinos, annihilation through the Higgs or $Z$ boson exchange  is expected to yield significant rates for direct detection.
The interaction of DM with nuclear matter can be classified as elastic or inelastic; and as spin-dependent or spin-independent.

In elastic scattering the DM interacts with the nucleus as a whole, causing the nucleus to recoil, while in inelastic scattering some of the energy goes into recoil, and some is used to excite the nucleus to a higher energy state, from where it decays by emitting a photon. The dark matter detection experiments (DAMA/LIBRA \cite{Bernabei:2010mq}, CoGeNT \cite{Aalseth:2010vx} and CRESST-II \cite{Angloher:2014myn})  have reported signals consistent with a light DM candidate and with an elastic cross section with nucleons of ${\cal O}(10^{-41}-10^{-40}{\rm cm^2})$.

In spin-dependent (axial vector) scattering, the DM spin couples with the spin of the nucleon, while in spin-independent (scalar) scattering, the cross section does not depend on this,  and thus  it is larger for larger nuclei because of the coherence of DM interacting with the nucleus as a whole. We analyze the predictions of our model for the spin-dependent and spin-independent cross sections in turn, and compare them with the experimental predictions.

In Fig. \ref{fig:SDlimit}, in the upper panels, we present the spin-dependent (SD) cross section of dark matter scattering off nucleons, as a function of the dark matter mass $M_{DM}$ for $\sin \alpha=0$. The left panel is for the proton, the right one for the neutron. The red lines show points of the parameter space, with restricted $M_{DM}$ and $h^\prime_{\nu}$ values, which reproduce acceptable relic density. The areas above the pink dashed line and green dashed-dotted line are ruled out by the COUPP \cite{Behnke:2012ys} and XENON100  \cite{Aprile:2013doa} measurements, respectively. As the plots show, to obtain the correct relic density, the resonantly enhanced annihilation rate implies a suppressed Yukawa coupling for the neutrino DM, which leads to a suppressed cross section. Here again we observe the two dips surrounding the $Z$ resonance and the $h$ resonance. The limits on the SD cross section from  COUPP and XENON100 results do not restrict the parameter space of our model.  In the bottom panel, we plot contour graphs for the spin-dependent cross sections of the nucleon as functions of the dark matter mass $M_{DM}$ and Yukawa coupling $h_\nu^\prime$, for $\sin \alpha=0$. Again we show the spin-dependent cross section of the proton and neutron in the left and right panel, respectively. All points are consistent with experimental bounds on the spin-dependent nucleon cross sections, as indicated by the color-coded panels, but only parameter points situated along the dashed-dotted lines in the bottom panels give the correct dark matter relic density. 
These cross sections are not sensitive to variations in $\sin \alpha$.

\begin{figure}[htbp]
\vskip -0.3in
\begin{center}$
    \begin{array}{cc}
    \includegraphics[width=3.3in,height=3in]{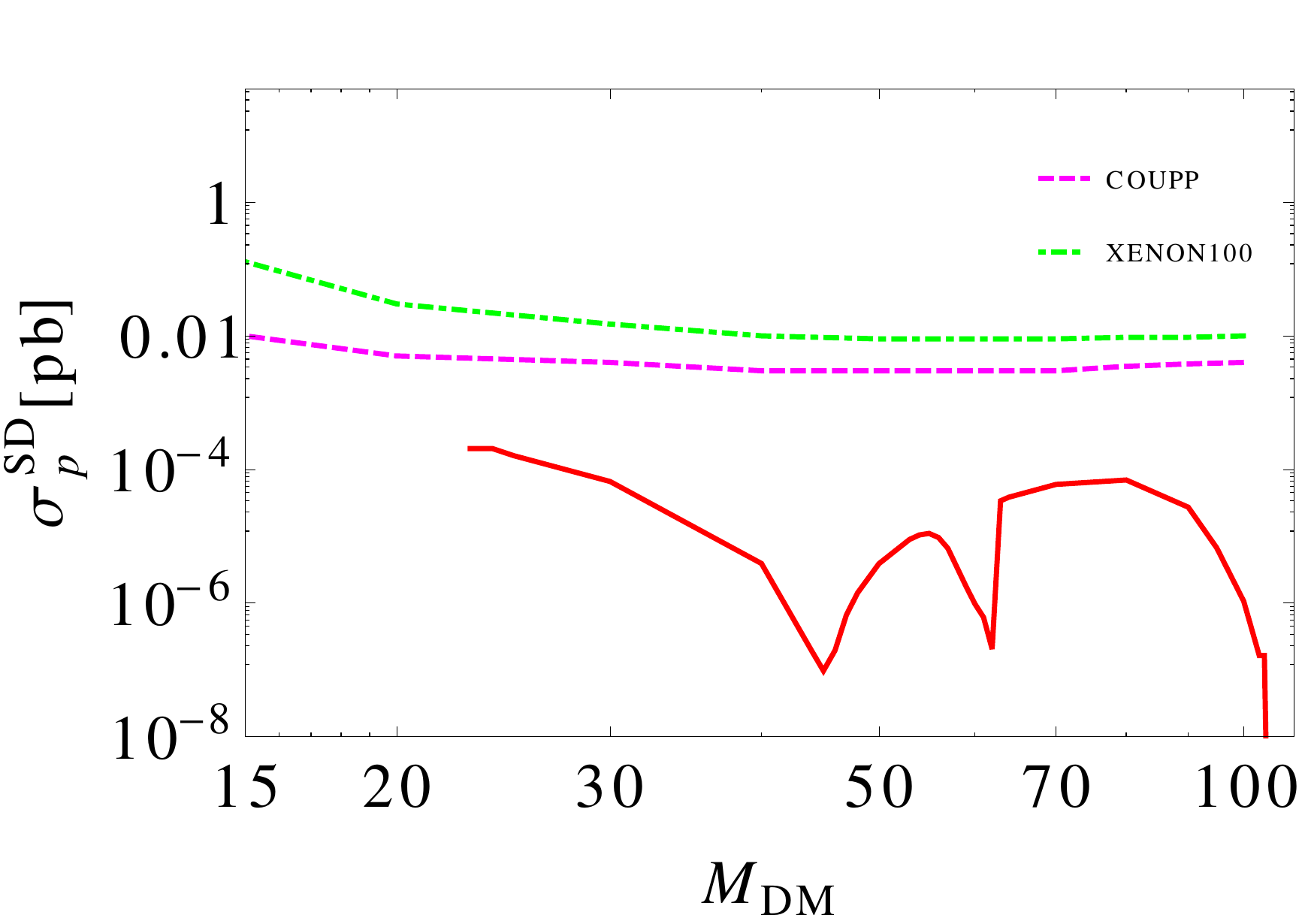}
&\hspace*{-0.2cm}
    \includegraphics[width=3.3in,height=3in]{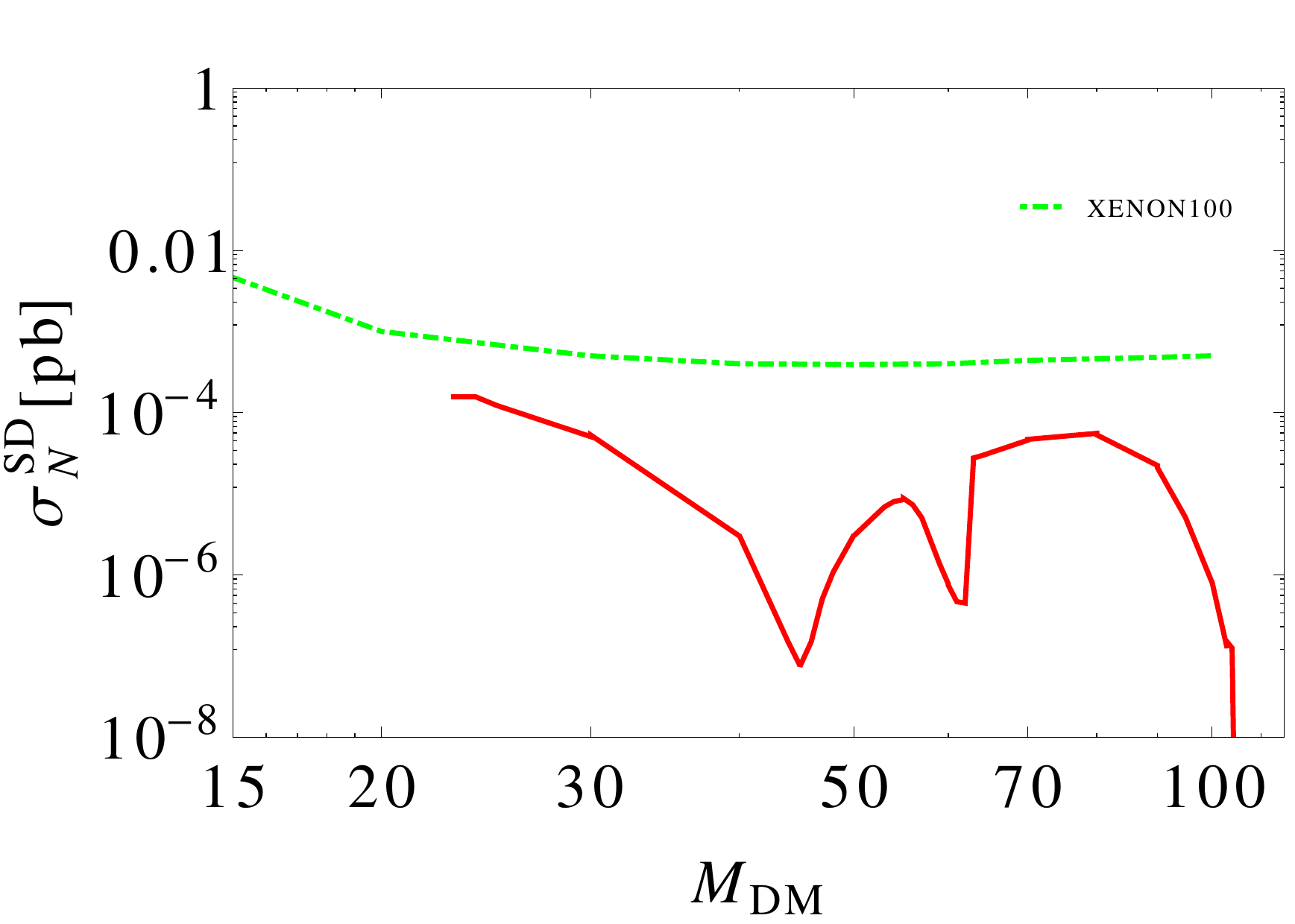}\\
        \includegraphics[width=3.3in,height=2.8in]{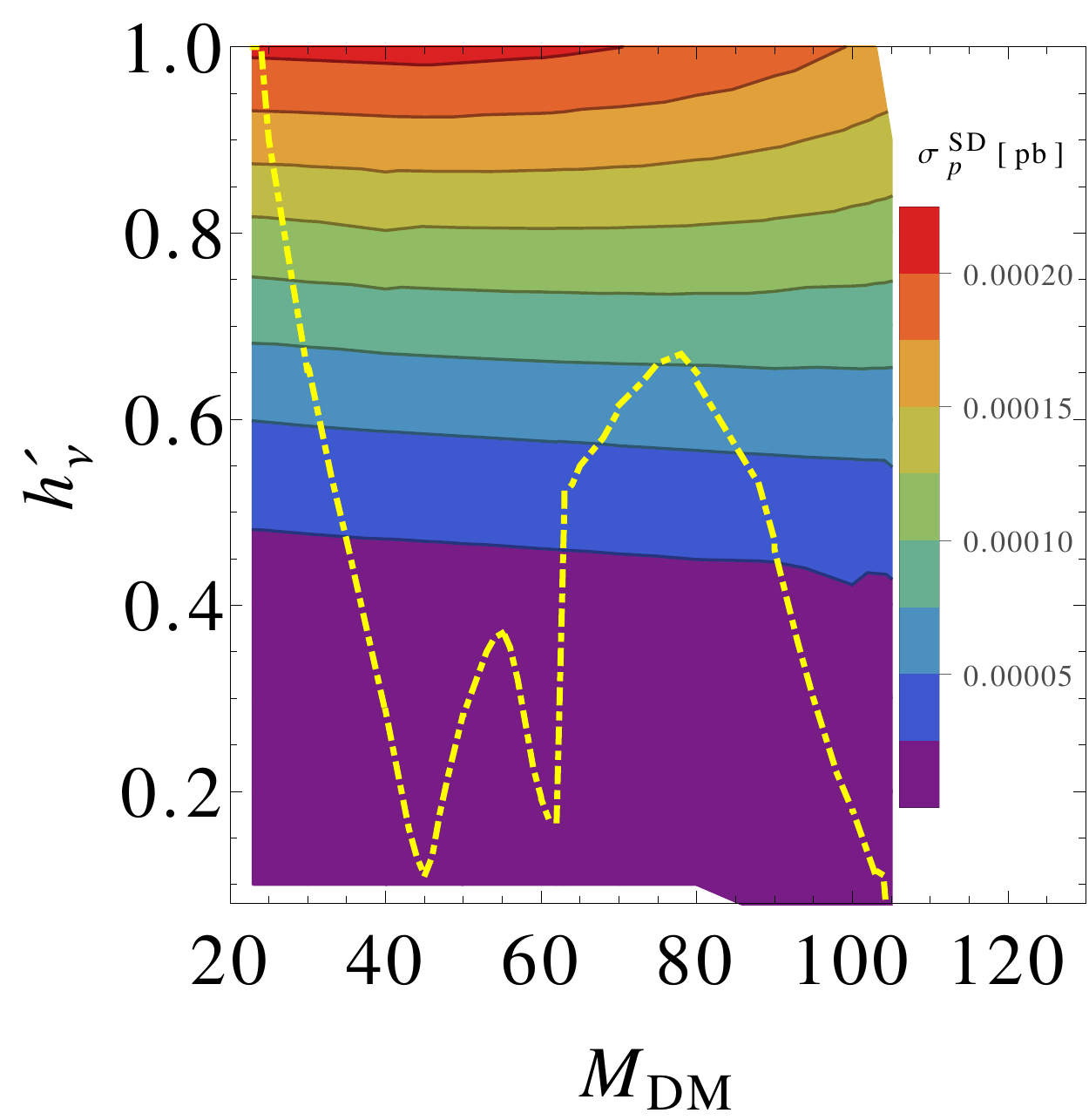}
&\hspace*{-0.2cm}
    \includegraphics[width=3.3in,height=2.8in]{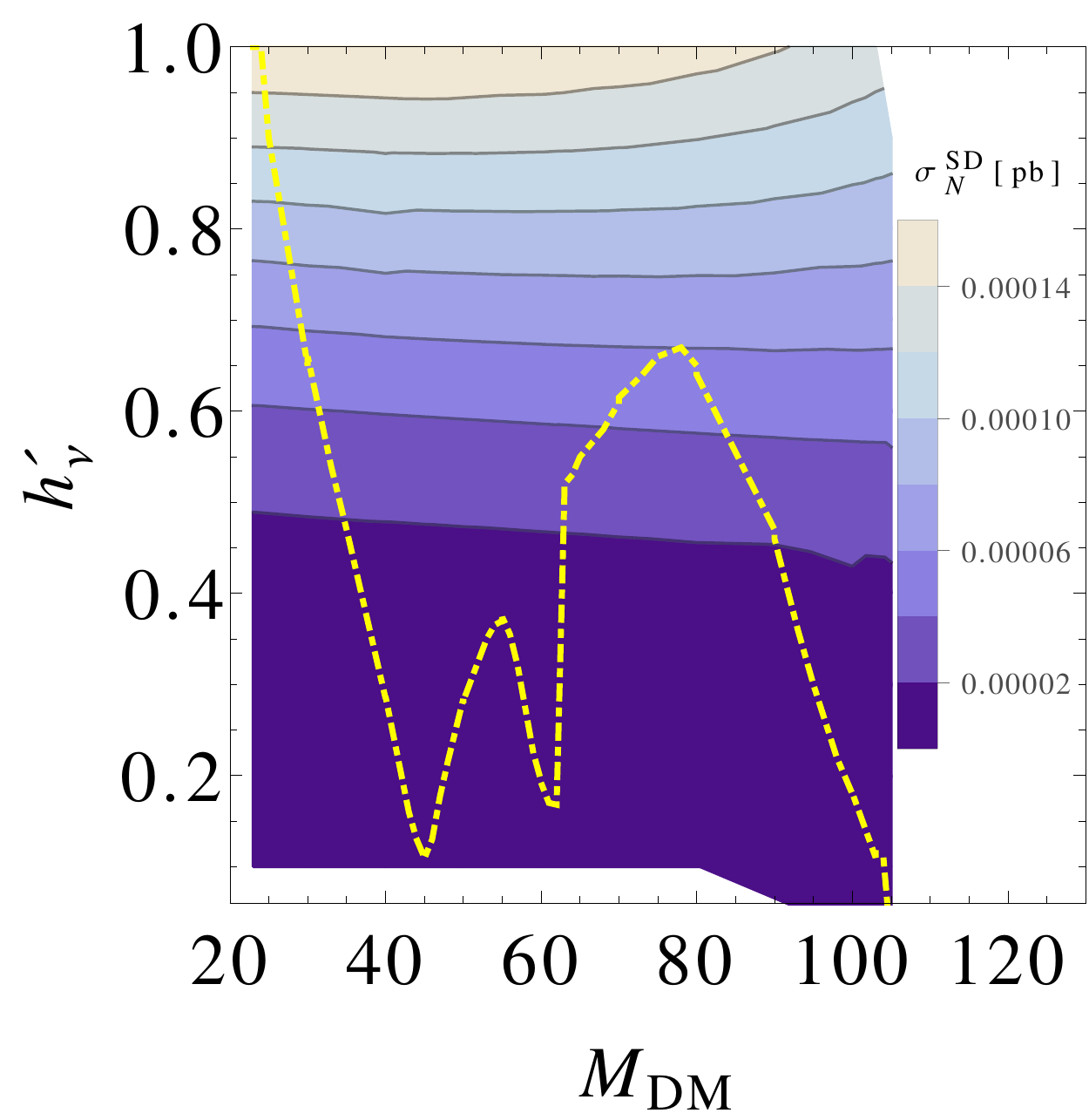}
        \end{array}$
\end{center}
\vskip -0.1in
     \caption{(color online). \sl\small Top: the spin-dependent cross section of the nucleon in the HTM with vector-like leptons, as a function of the dark matter mass $M_{DM}$ (GeV) for $\sin \alpha=0$.  We show (left panel) the spin-dependent cross section of the proton (red line) with XENON100 (dash-dotted green) and COUPP (dashed pink) \cite{Aprile:2013doa} results,  (right panel) the spin-dependent cross section of the neutron (red line) with XENON100 (dash-dotted green) results \cite{Aprile:2013doa}. The area above the pink dashed line and green dash-dotted line are ruled out by the COUPP and XENON100  results respectively.  Bottom: Contour plots showing the spin-dependent cross section of the nucleons in the HTM with vectorlike leptons, as functions of the dark matter mass $M_{DM}$ and Yukawa coupling $h_\nu^\prime$, for $\sin \alpha=0$. We show the spin dependent cross section of the proton (left panel) and the spin-dependent cross section of the neutron ( right panel). The panels at the right indicate the color-coded values of the cross section along each slice, and the dashed-dotted line represents the only parameter  points with acceptable relic density,  $0.1144 \leq \Omega_{DM} h^2 \leq 0.1252$. 
     }
\label{fig:SDlimit}
\end{figure}

In Fig. \ref{fig:SI}, we plot the spin-independent (SI) cross section of nucleon, as a function of the dark matter mass $M_{DM}$ (in GeV) for $\sin \alpha=0$ (left  panel). The red line includes all points yielding consistent relic density. The regions above dash-dotted black line, dash-dotted green line, dash-dotted orange line, dash-dotted blue line, dash-dotted purple line, dash-dotted pink line are ruled out by XENON100 \cite{Lavina:2013zxa}, XENON100 with $2\sigma$ expected sensitivity, CRESST-II \cite{Angloher:2014myn}, CDMS-II \cite{Agnese:2013cvt}, TEXONO \cite{Li:2013fla} and DAMIC100 (expected for 2014)  \cite{Chavarria:2014ika} results, respectively. The cross section is enhanced at the $Z$ pole and $h$ pole and there, for a suppressed direct rate, the  Yukawa coupling must be suppressed to compensate for the resonant production effect. This is seen as two dips at $M_{DM}\sim M_Z/2$ and $M_{DM}\sim m_h/2$.   The limit on the SI cross section from XENON100 constrains strongly our model, while the updated results from the other experimental results do not restrict the parameter space. As the left panel of the figure shows, XENON100 results (with $2\sigma$ expected sensitivity) restrict the dark matter mass to be in the  37-52 GeV, or  57-63 GeV ranges, or heavier than 95 GeV. In the middle panel, we show the spin-independent cross section of the proton as a graph in $M_{DM}-h_\nu^\prime$ space,  constrained by all the experiments with the exception of XENON100, while in the right panel we include XENON100 measurements. The latter rules out large regions of parameter space  (in white) while in both panels colored contours (as coded in the attached bars) are allowed by the spin-dependent experiments. In both middle and right panels, the dashed-dotted line represents the only parameter points with acceptable relic density. Note here that, in agreement with the left panel, there are regions of the parameter space where no combination of $M_{DM}$ and $h^\prime_{\nu}$ satisfy both relic density and XENON100 SI cross section restrictions. Here too, the cross sections are not sensitive to the mixing angle or to other parameters in the model, and for the spin-independent, the cross sections for the proton and neutron are indistinguishable.

\begin{figure}[htbp]
\vskip -0.3in
\begin{center}$
    \begin{array}{ccc}
	\hspace*{-1.1cm}\includegraphics[width=2.5in,height=3.0in]{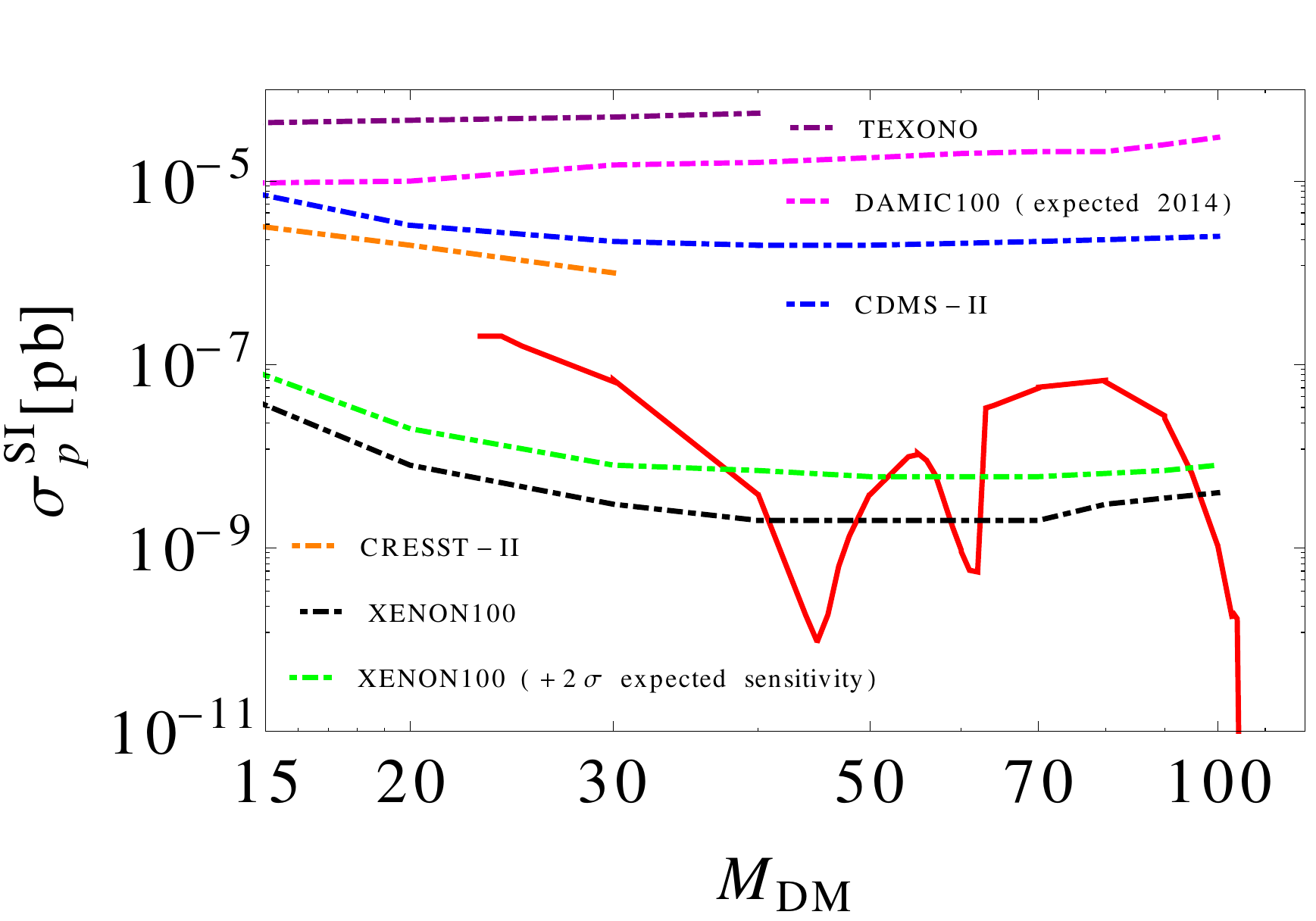}
	& \hspace*{-0.2cm}  \includegraphics[width=2.5in,height=2.8in]{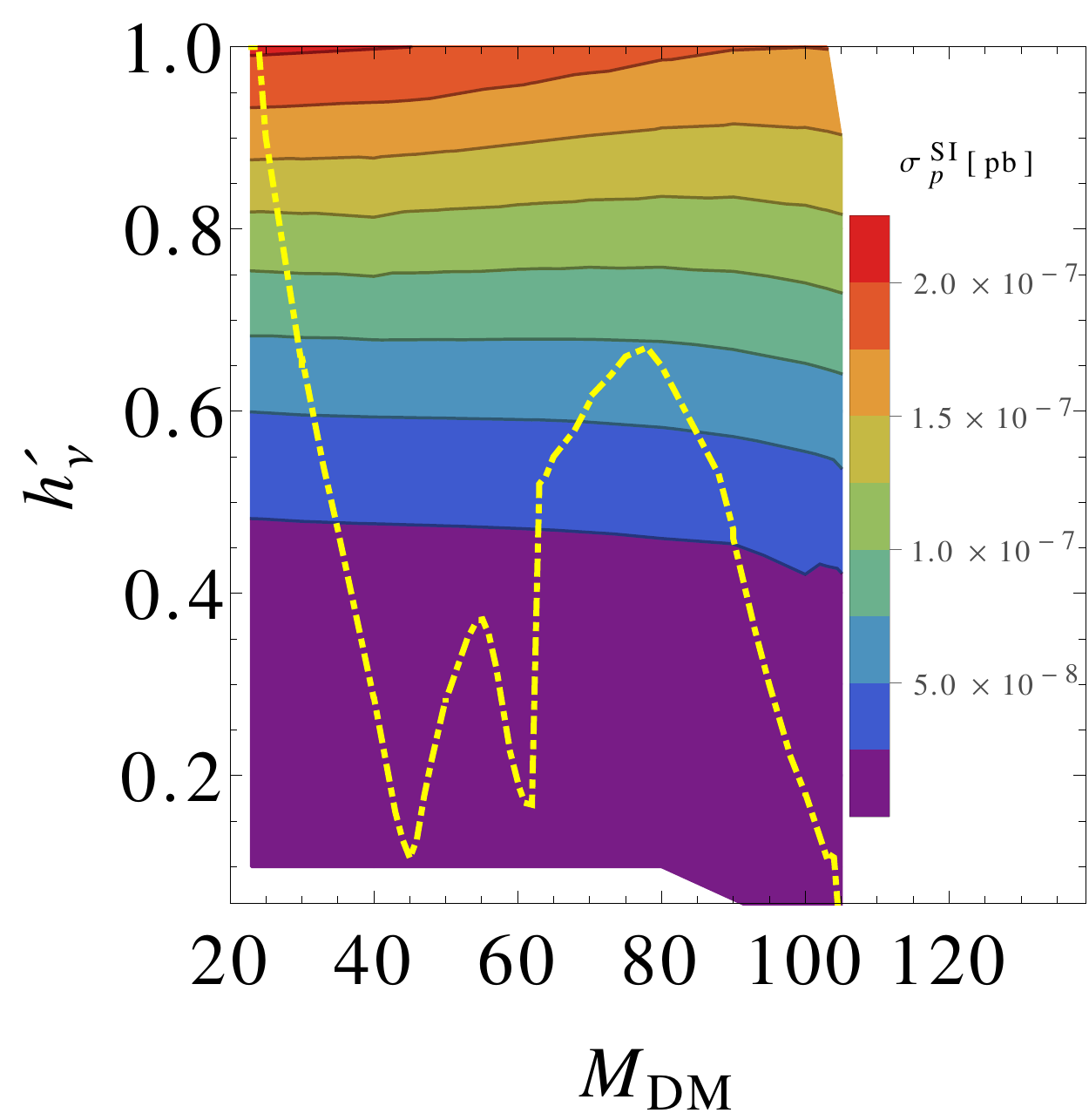}
	& \hspace*{-0.2cm}  \includegraphics[width=2.4in,height=2.8in]{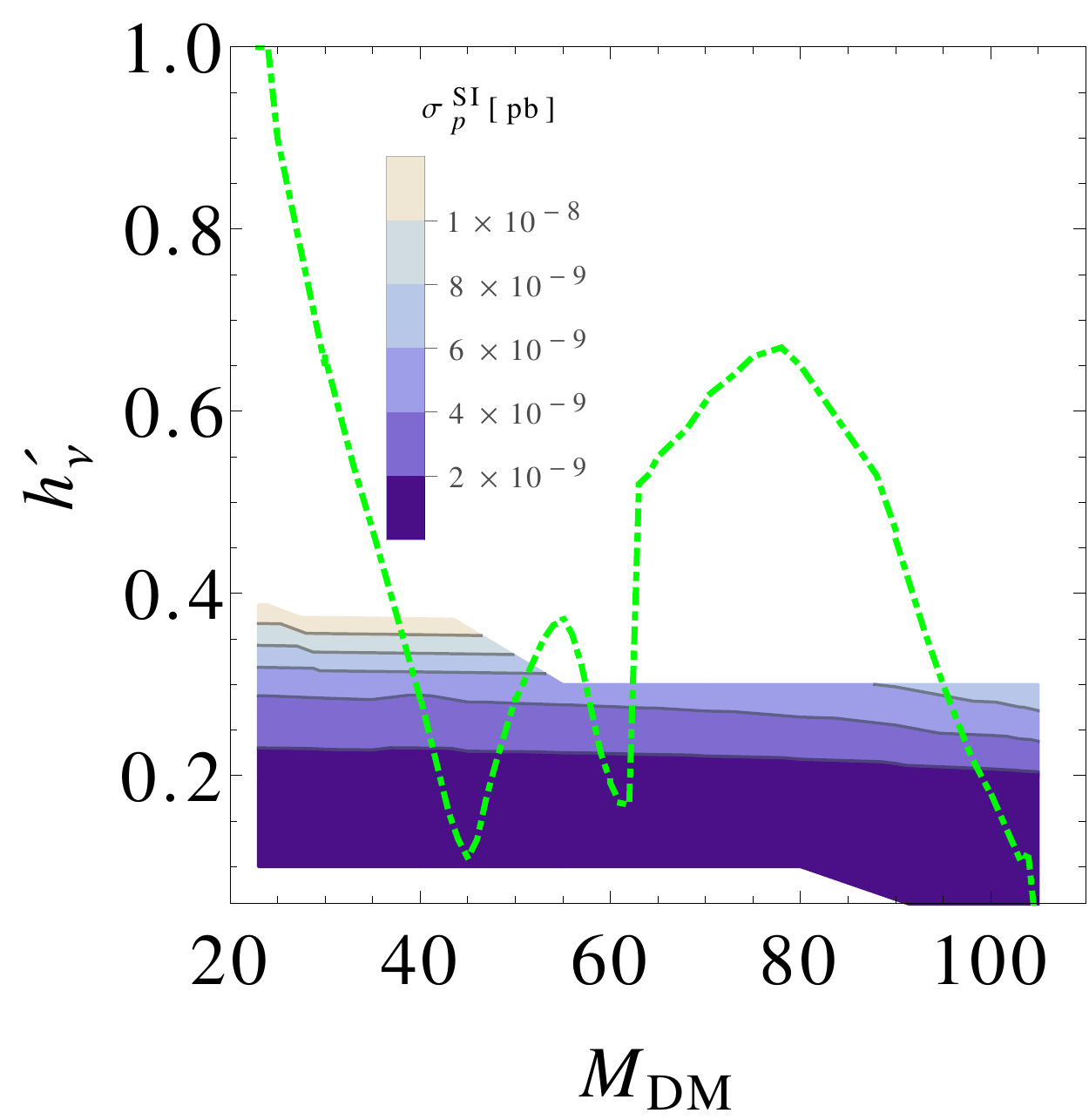}
        \end{array}$
\end{center}
\vskip -0.1in
     \caption{(color online). \sl\small Left panel: the spin-independent cross section of the proton as a function of the dark matter mass $M_{DM}$ (GeV) in the HTM (red line). We also show XENON100 \cite{Lavina:2013zxa} (dash-dotted black) XENON100 with $2\sigma$ expected sensitivity (dash-dotted green), CRESST-II \cite{Angloher:2014myn} (dash-dotted orange), CDMS-II \cite{Agnese:2013cvt} (dash-dotted blue), TEXONO \cite{Li:2013fla} (dash-dotted purple) and DAMIC100 (expected in 2014) \cite{Chavarria:2014ika} (dash-dotted pink) results.
Middle panel: Contour graph showing the spin independent cross sections of nucleon in the HTM with vector-like leptons, as functions of the dark matter mass $M_{DM}$ and $h_\nu^\prime$ for for $\sin \alpha=0$,  considering all experimental constraints except XENON100. Right panel: same as the middle panel, but including constraints for XENON100 (with $+2\sigma$ expected sensitivity) \cite{Lavina:2013zxa} upper limit. The values of the cross  section are indicated on the plots. The contours indicate points consistent with the respective experimental constraints, while the dashed-dotted line includes only points with acceptable relic density, $0.1144 \leq \Omega_{DM} h^2 \leq 0.1252$.  }
\label{fig:SI}
\end{figure}

\section{Indirect Detection}
\label{sec:ID}
%

  Pairs of dark matter particles annihilate producing high-energy particles (antimatter, neutrinos or photons). Indirect detection experiments for dark matter look for signatures of annihilations of DM originating from particles in the flux of cosmic rays and are sensitive to DM interaction with all the SM particles. The most stringent constraints on DM annihilation cross sections have been derived from the Fermi Gamma Ray Space telescope (Fermi-LAT) \cite{Ackermann:2013yva},  used to search for DM annihilation products from dwarf spheroidal galaxies and the Galactic Center, which probe annihilation cross sections into photons of $\langle \sigma v\rangle \sim 3 \times 10^{-26} {\rm cm^3/s}$. 
 These searches have  attracted a lot of attention due to the unexpected high flux of cosmic ray positrons observed by the PAMELA experiment \cite{Adriani:2008zr}, and confirmed by AMS  \cite{Accardo:2014lma}.

In Fig. \ref{annihilation}, we present the annihilation cross section of DM as a function of the dark matter mass $M_{DM}$ and compare it with the constraints on the dark matter annihilation cross section for the $e^+ e^-$ channel, $\mu^+ \mu^-$ channel, $\tau^+ \tau^-$ channel, $u\bar{u}$, $b\bar{b}$ channel and $W^+ W^-$ channel at 95\% CL, derived from a combined analysis of 15 dwarf spheroidal galaxies of Fermi-LAT Collaboration results \cite{Ackermann:2013yva} (left panel). As the figure shows, the limit on the annihilation cross section  from Fermi-LAT Collaboration results does not impose any restriction on our model parameters.  Again, the annihilation cross section is enhanced at the $Z$ pole around $M_{DM}=M_Z/2$. 
The regions around $M_{DM}=M_Z/2$ can be brought into agreement with the relic density constraint by modifying the neutral Yukawa coupling $h_\nu^\prime$. In order to treat the regions where the annihilation cross section is enhanced, we need to decrease the value of Yukawa coupling. A suppressed coupling leads to suppression of the annihilation rates \cite{Joglekar:2012vc, Heikinheimo:2012yd}.   The effect of the Higgs pole at $M_{DM}\sim h/2$ is more dramatic than the effect of the $Z$ pole. The dominant annihilation modes of dark matter pair in this region are coming from decay into quark/antiquark (mainly $b$ $\bar b$,   which gives a relative contribution of $\sim 77\%$ to $1/\Omega_{DM} h^2$) and also small contribution from $c\bar c$ and $\tau \bar \tau$ to obtain the correct dark matter relic density. In the right panel, we show the annihilation cross section  as a contour plot in the dark matter mass $M_{DM}$ and Yukawa coupling $h_\nu^\prime$ plane. Here, as in the previous figures, the contours, according to the color-coding in the attached bar, represent the regions of the parameter space consistent with the experimental results, while the white regions are excluded. Only points along the dashed-dotted line have acceptable relic density.   
\begin{figure}[htbp]
\vskip -0.3in
\begin{center}$
    \begin{array}{cc}
\hspace*{-0.5cm}
\includegraphics[width=3.3in,height=3in]{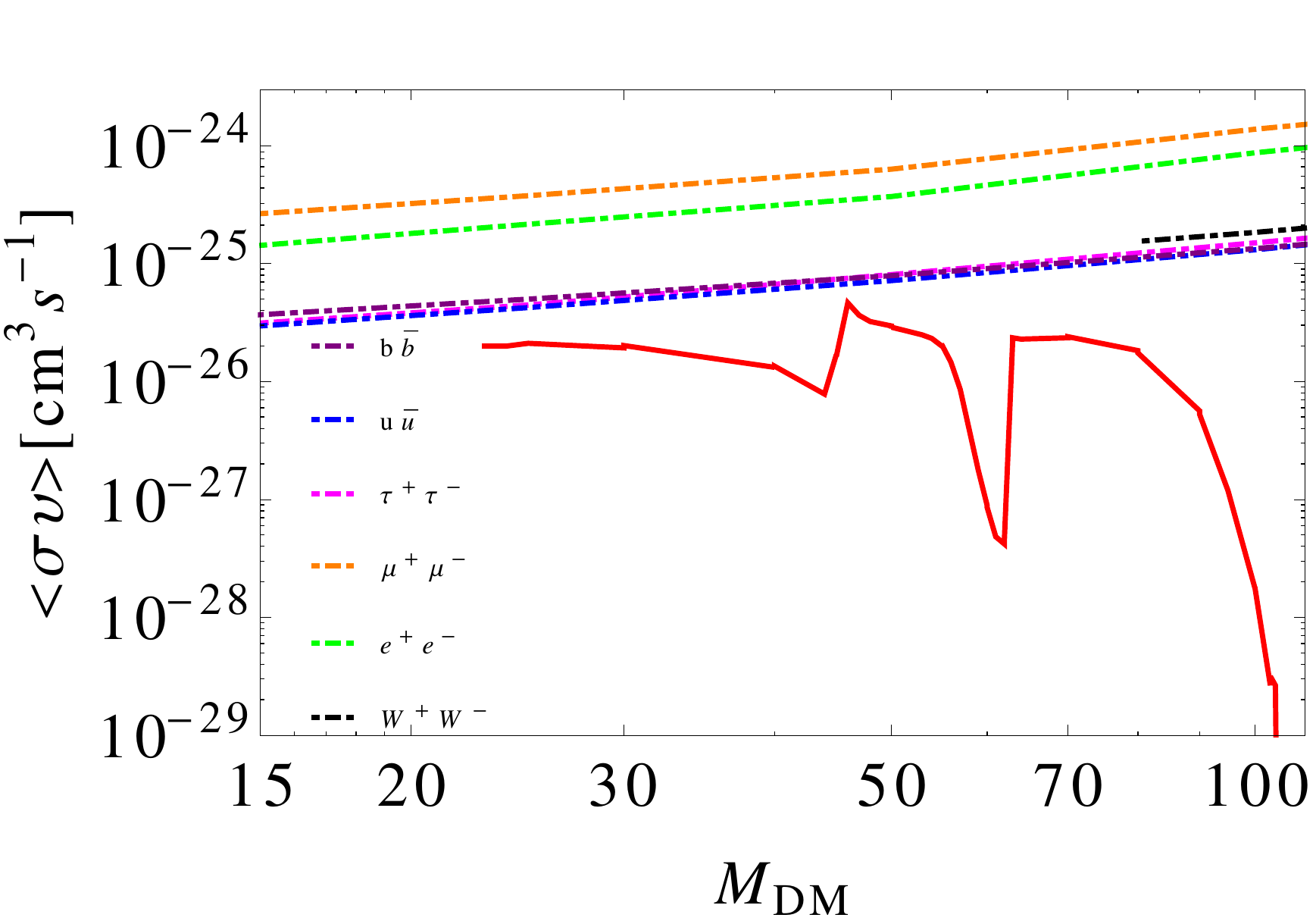}
    &\hspace*{-0.2cm}
   \includegraphics[width=3.4in,height=2.8in]{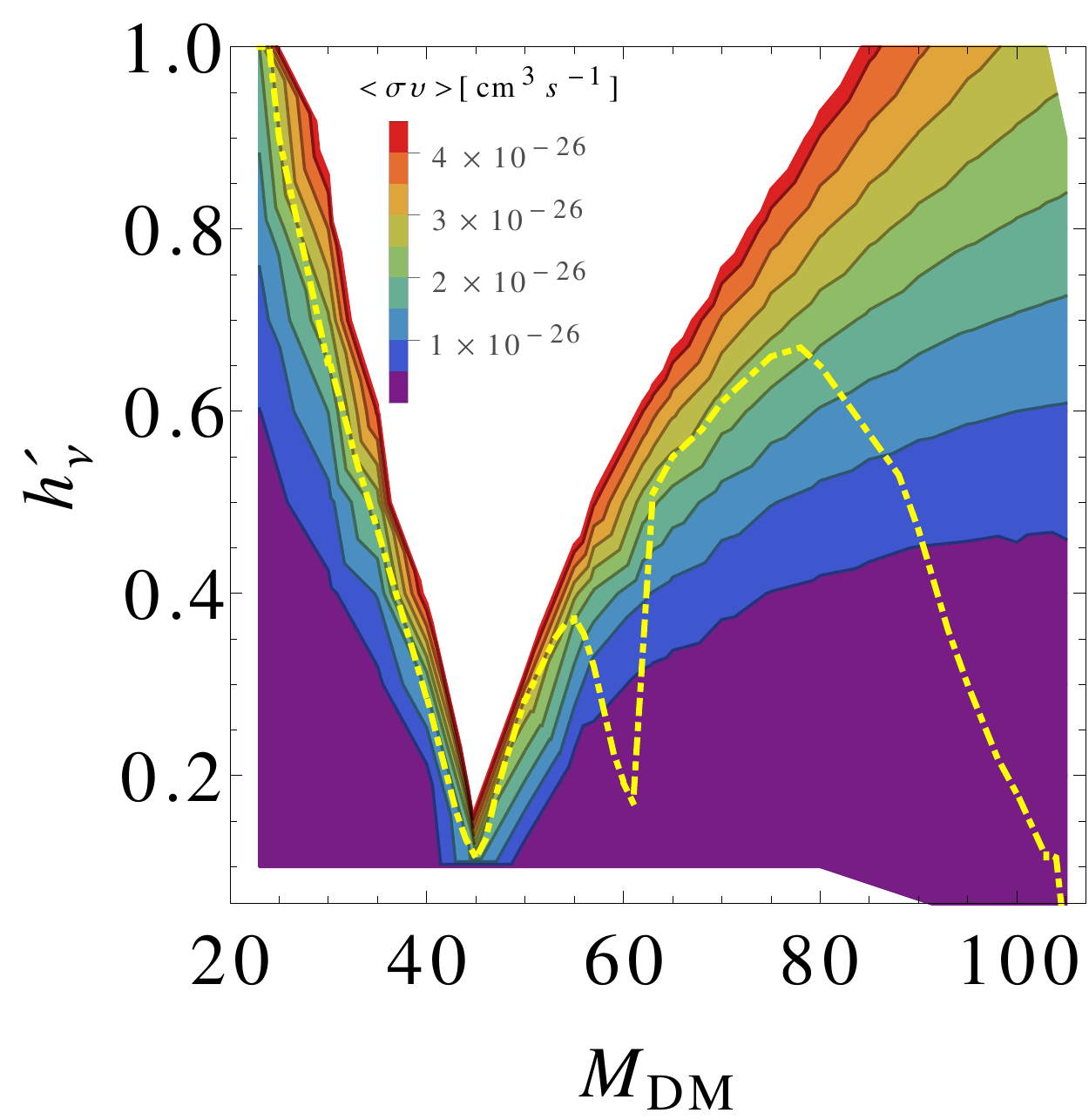}
  \end{array}$
\end{center}
\vskip -0.1in\caption{(color online). \sl\small Left panel: the annihilation cross section of  DM  as a function of the dark matter mass $M_{DM}$ (GeV) (red line).  We show the constraints on the dark matter annihilation cross section for the $e^+ e^-$ channel (dash-dotted green), $\mu^+ \mu^-$ channel (dash-dotted orange), $\tau^+ \tau^-$ channel (dash-dotted pink), $u\bar{u}$ (dash-dotted blue), $b\bar{b}$ channel (dash-dotted purple) and $W^+ W^-$ channel (dash-dotted black)  at 95\% CL derived from the combined analysis from Fermi-LAT Collaboration  \cite{Ackermann:2013yva}.  Right panel: contour plot showing the annihilation cross section  as function of the dark matter mass $M_{DM}$ and its Yukawa coupling $h_\nu^\prime$. The contours are consisted with the experimental values for the cross sections, indicted by the color-coded bar, while the white regions are ruled out. Only  points along the dashed-dotted  line give the correct dark matter relic density, $0.1144 \leq \Omega_{DM} h^2 \leq 0.1252$. }
\label{annihilation}
\end{figure}

\section{Detection at Particle Colliders}
\label{sec:LHC}

If dark matter has significant coupling to nuclear matter it can be produced in high energy collisions at LHC or at future colliders. Once produced, as it is neutral and weakly interacting, DM will not be observed directly, but it could inferred from missing transverse momentum. Collider searches provide the opportunity to study DM production in a controlled environment. They are particularly sensitive to the region of low mass dark matter, where backgrounds are smaller. At the LHC, dark matter can be produced directly, together with additional radiation from the quarks or gluons participating in the reaction, which results in a single jet (mono-jet) plus missing momentum. High energy lepton colliders could create dark matter through a similar process. Assuming DM couples to quarks and gluons and couplings the order of the electroweak size, LHC excludes DM masses up to 500 GeV and for DM coupling to electrons with the same-size couplings, LEP excludes DM with mass below 90 GeV. Neither of these restrictions are applicable here, as vectorlike neutrinos do not couple directly to either quarks or leptons. 

\section{The Flux of Muons and Neutrinos from the Sun}
\label{sec:NMF}
%

The recent observation of ultra-high energy 
neutrino events at IceCube \cite{Aartsen:2013bka} seem to indicate a possible deficit in the muon track (known as the muon deficit problem) and an apparent energy gap in  the 3-year  high energy neutrino data, challenging a simple explanation in terms of atmospheric neutrinos and suggesting an extra-terrestrial
origin. These astrophysical neutrinos are assumed to have originated from the decays of charged particles produced in $pp$ or $p \gamma$ collisions. While the data obtained is largely consistent with SM predictions, the flux shows a mild deficiency in muons at high energies, prompting alternative explanations involving dark matter.

In Fig. \ref{fig:NMfluxlimit}, we show the neutrino (left panel) and muon (right panel) fluxes as functions of the dark matter mass $M_{DM}$ (GeV). On the top graphs, we plot our results as a red curve, and include the upper limits on the neutrino and muon flux for the $b\bar{b}$ channel, $\tau^+ \tau^-$ channel, and the $\nu_e \bar{\nu_e}$, $\nu_{\mu} \bar \nu_{\mu}$,  $\nu_{\tau} \bar \nu_{\tau}$ channels from the Baikal NT200 detector results \cite{Avrorin:2014swy}. While the limit on the muon flux from Bailkal results does not impose  restrictions on our model\footnote{Due to limited space on the figure, we show the recent results of Baikal NT200. But our results are also consistent with those from the Baksan
Neutrino Observatory \cite{Boliev:2013ai}.},  the neutrino flux excludes DM particles with mass in the 74-85 GeV.  The figures show again the two dips at $M_{DM}\sim 45$ GeV and $M_{DM}\sim 62$ GeV. Unlike in the annihilation cross section, here the effect of the $Z$ pole is more dramatic than that of the Higgs pole at $M_{DM}\sim h/2$. The bottom panels show the fluxes of neutrino and muon as contour plots in the dark matter mass $M_{DM}$ and the Yukawa coupling $h_\nu^\prime$. Note that here, as before the contours are consistent with the experimental values for the measured flux of muons and neutrinos.  However only points along the dashed-dotted line are consistent with the dark matter relic density exclusion limit.


\begin{figure}[htbp]
\vskip -0.3in
\begin{center}$
    \begin{array}{cc}
    \includegraphics[width=3.4in,height=3in]{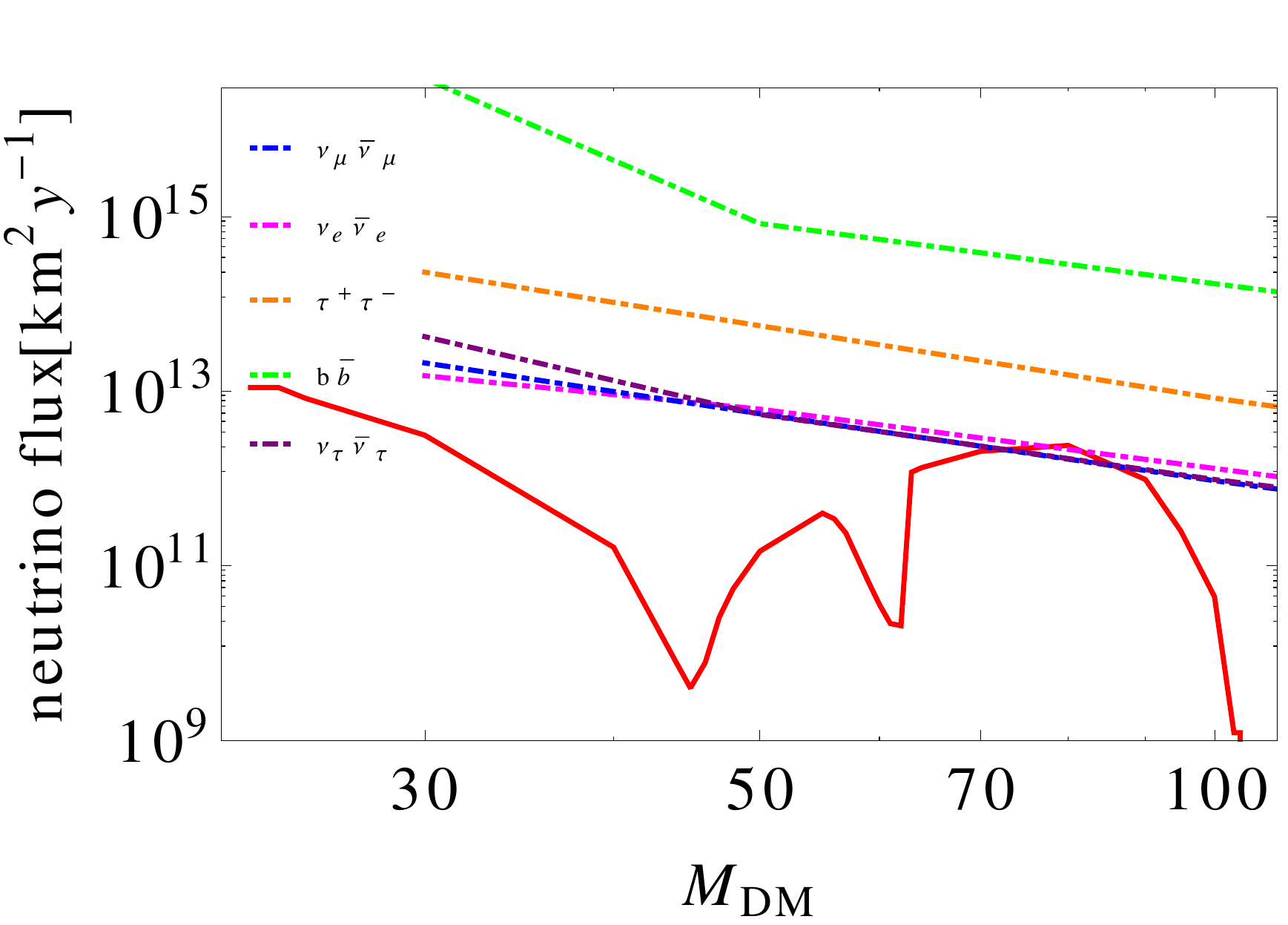}
&\hspace*{-0.1cm}
    \includegraphics[width=3.4in,height=3in]{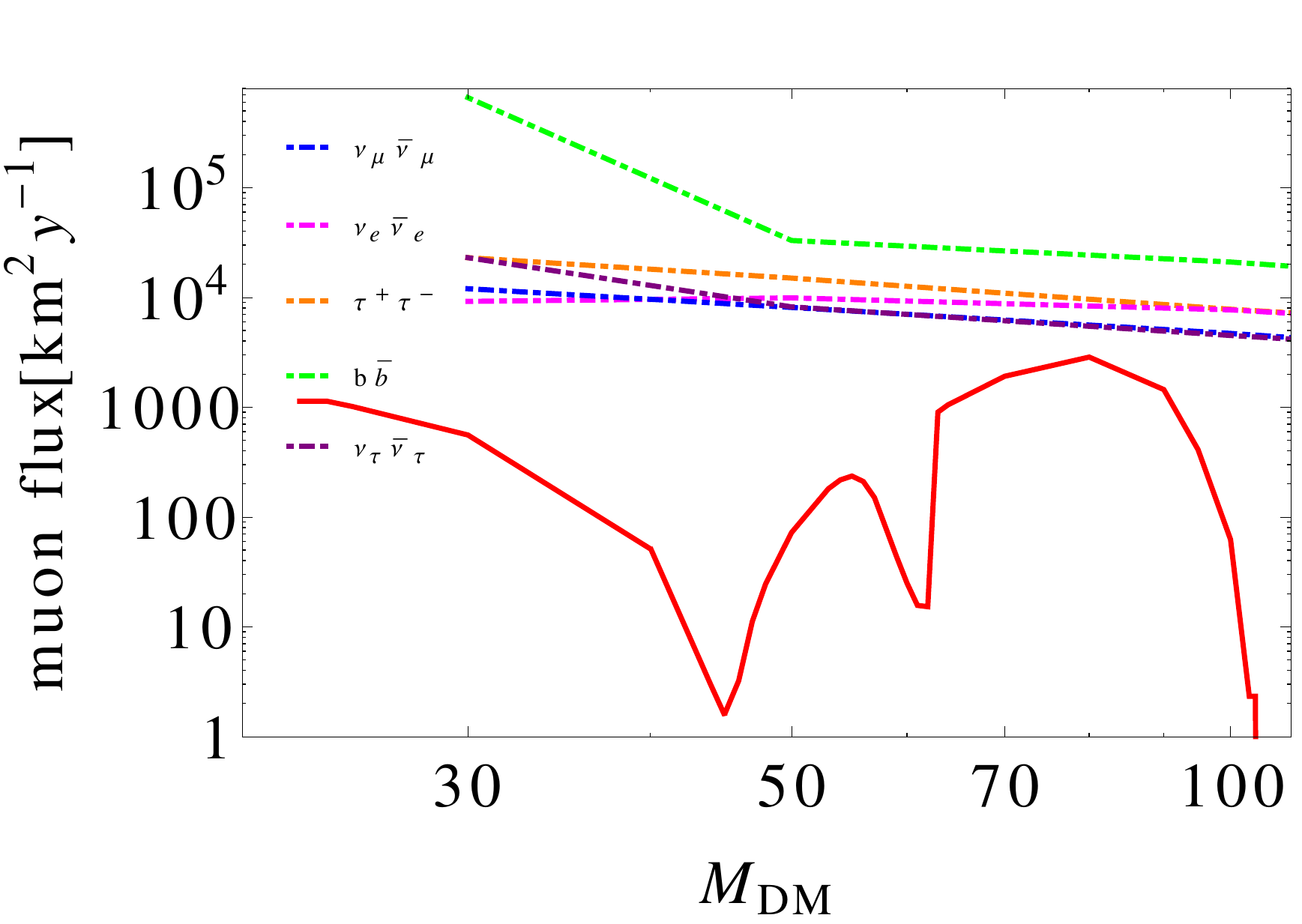}\\
           \includegraphics[width=3.4in,height=2.8in]{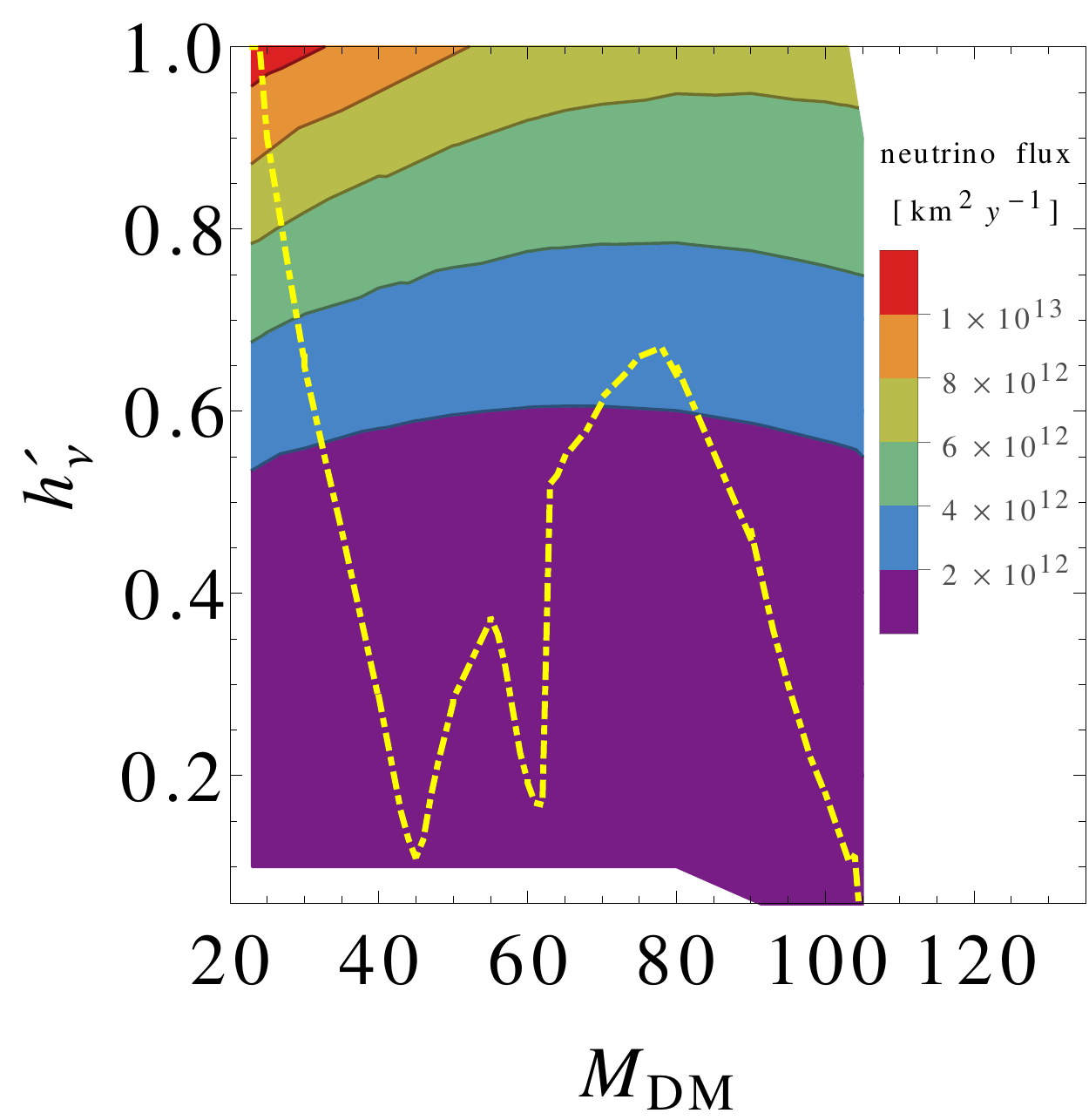}
&\hspace*{-0.2cm}
    \includegraphics[width=3.4in,height=2.8in]{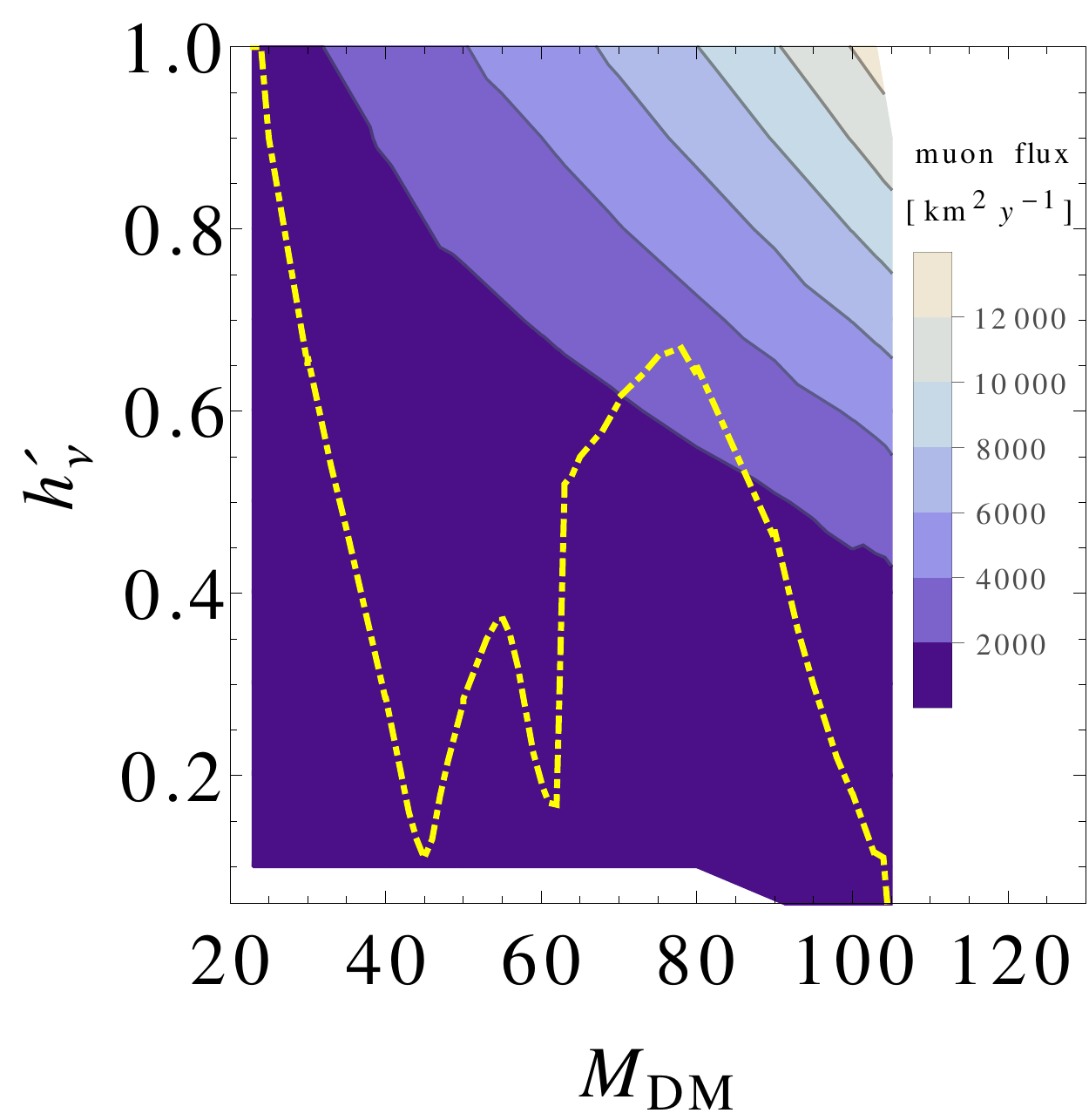}
        \end{array}$
\end{center}
\vskip -0.1in
     \caption{(color online). \sl\small Top: the flux of neutrinos (left panel) and muons (right panel) in the HTM with vectorlike leptons, as functions of the dark matter mass $M_{DM}$ (GeV) (red line). We also show the upper limits on the neutrino and muon flux for the $b\bar{b}$ channel (dash-dotted green), $\tau^+ \tau^-$ channel (dash-dotted orange), $\nu_e \bar{\nu_e}$ channel (dash-dotted pink), $\nu_{\mu} \bar \nu_{\mu}$ channel (dash-dotted blue) and $\nu_{\tau} \bar \nu_{\tau}$ channel (dash-dotted purple) from the Baikal NT200 detector results \cite{Avrorin:2014swy}. Bottom: contour plots showing the flux of neutrinos (left panel) and muons (right panel) in the HTM with vectorlike leptons, as functions of the dark matter mass $M_{DM}$ and Yukawa coupling $h_\nu^\prime$. The flux values are color-coded as in the bar attached. The contours indicate points consistent with the respective experimental constraints, while the dashed-dotted line includes the only points with acceptable relic density, $0.1144 \leq \Omega_{DM} h^2 \leq 0.1252$.  }
\label{fig:NMfluxlimit}
\end{figure}



\clearpage

\section{Summary and Conclusion}
\label{sec:conclusion}

In this work we analyzed the effects of introducing vectorlike leptons in the Higgs Triplet Model. Our aim was to provide a scenario that can explain both neutrino masses and provide a DM candidate, problems unresolved in the SM. We choose  a full generation of vectorlike leptons (one left-handed doublet, two right-handed singlets, together with their mirror representations). We insure that a new symmetry differentiates between ordinary leptons and the new states, forbidding unwanted lepton flavor violation. Opting for a simplified Yukawa coupling structure, a mostly singlet right-handed vectorlike neutrino emerges as a single DM candidate. 
Introducing vectorlike leptons in the HTM relaxes the severe constraints on the mass of the doubly charged Higgs boson coming from electroweak precision tests. We revisit precision observables in this work, and show that while the $S$ parameter does not impose constraints on the parameter space, the $T$ parameter is restrictive, allowing only certain combinations of doubly-charged mass, Yukawa couplings and mixing angles between the neutral Higgs bosons. Of these, the most sensitive parameter is the mass of the doubly charged Higgs boson, required to be less than about 280 GeV, but this boson has different branching ratios than in the minimal HTM. The $T$ parameter is insensitive to the mass of the dark matter candidate.
 
We verify that the invisible decay width of the Higgs boson is consistent not only with the experimental data, but with the more restrictive limits imposed by global fits to the Higgs data. The invisible width is a relevant constraint for dark matter masses less than 1/2 the Higgs mass, and all of these survive. More stringent constraints come from direct detection experiments, especially from restriction on spin-independent nucleon cross section, and from the relic density.
The latter restricts the combination between dark matter mass and its Yukawa coupling to narrow bands in the parameter space, and disallows entirely regions where the DM candidate is lighter than 23 GeV, or heavier than 103 GeV. If one includes constraints from XENON100 on spin-independent scattering of dark matter off nucleons, these further restrict the dark matter mass to be in the the ranges: 37-52 GeV, or 57-63 GeV, or heavier than 95 GeV, all for points satisfying relic density constraints.  In addition, consistent with direct detection experiments,  the neutrino flux excludes DM particles with mass in the 74-85 GeV range. These are the most stringent restrictions, and they are insensitive to other model parameters, such as other masses (particularly the doubly charged Higgs boson) and the mixing angle between the neutral Higgs bosons.

To summarize, we have presented a simple model that accounts for neutrino masses and dark matter and is consistent with the relic density and all direct and indirect searches. This model assumes a single dark matter particle, and the experimental data restricts its mass to be confined to limited regions in the parameter space.  If the dark matter is as light as keV or a few GeV, as some experiments suggest, this scenario is ruled out. However for DM mass around 30 GeV, allowing small deviations from direct detection, the HTM with vectorlike leptons provides a viable explanation. This analysis assumed the DM candidate to be light and set an upper bound of 108 GeV, by the choice of the mass of the lightest verctorlike charged lepton. One can extend this scenario to a more complicated one, involving several DM particles. This model would less constrained, but it looses the predictability of the simple scenario presented here. 
Given the importance of DM in understanding the universe, and the effort going into direct and indirect detection, and into collider experiments, simple models such as this one can help elucidating the nature of DM.

\section{Acknowledgments}
We thank NSERC for partial financial support  under grant number
SAP105354.




\begin{thebibliography}{99}

\bibitem{LHCHiggs}
  G.~Aad {\it et al.}  [ATLAS Collaboration],
  Phys.\ Lett.\ B {\bf 716} (2012) 1;
  S.~Chatrchyan {\it et al.}  [CMS Collaboration],
  Phys.\ Lett.\ B {\bf 716} (2012) 30.
  
  \bibitem{pdg}
  K.A. Olive et al. [Particle Data Group Collaboration], Chin. Phys. C 38, 090001 (2014).


\bibitem{seesaw}
P. Minkowski, Phys. Lett. 67B, 421 (1977); T. Yanagida,
in Conf. Proc., C7902131, 95 (1979); Prog. Theor. Phys.
64, 1103 (1980); M. Gell-Mann, P. Ramond, and R. Slansky, in Supergravity edited by D. Z. Freedom and P.
van Nieuwenhuizen (North-Holland, Amsterdam, 1979);
  S.~L.~Glashow,
  NATO Sci.\ Ser.\ B {\bf 59}, 687 (1980).


\bibitem{seesaw2}
 R. N. Mohapatra and G. Senjanovic, Phys. Rev. Lett. 44,
912 (1980); 
  R.~N.~Mohapatra and G.~Senjanovic,
  Phys.\ Rev.\ D {\bf 23} (1981) 165.


\bibitem{htm}
W. Konetschny and W. Kummer, Phys. Lett. 70B, 433
(1977); M. Magg and C. Wetterich, Phys. Lett. 94B, 61
(1980); T. P. Cheng and L. F. Li, Phys. Rev. D 22, 2860
(1980); J. Schechter and J.W. F. Valle, Phys. Rev. D 22,
2227 (1980); G. Lazarides, Q. Shafi, and C. Wetterich,
Nucl. Phys. B181, 287 (1981).

  \bibitem{htm2}
A.~G. Akeroyd, Mayumi Aoki, and Hiroaki Sugiyama.
 Phys. Rev. D {\bf 77}, 075010  (2008);
A.~G. Akeroyd, Mayumi Aoki, and Hiroaki Sugiyama.
 Phys. Rev.  D {\bf 79}, 113010 (2009);
Takeshi Fukuyama, Hiroaki Sugiyama, and Koji Tsumura.
 JHEP {\bf 03}, 044 (2010);
S.~T. Petcov, H.~Sugiyama, and Y.~Takanishi.
 Phys. Rev. D {\bf 80}, 015005 (2009);
Takeshi Fukuyama, Hiroaki Sugiyama, and Koji Tsumura.
Phys. Rev. D {\bf 82}, 036004 (2010);
A.G. Akeroyd, Cheng-Wei Chiang, and Naveen Gaur,
JHEP {\bf 1011}, 005 (2010);
  A.~Arhrib, R.~Benbrik, M.~Chabab, G.~Moultaka and L.~Rahili,
  JHEP {\bf 1204}, 136 (2012);
  A.~Arhrib, R.~Benbrik, M.~Chabab, G.~Moultaka and L.~Rahili,
  [arXiv:1202.6621 [hep-ph]];
  A.~Arhrib, R.~Benbrik, M.~Chabab, G.~Moultaka, M.~C.~Peyranere, L.~Rahili and J.~Ramadan,
  Phys.\ Rev.\ D {\bf 84}, 095005 (2011);
E.~J.~Chun, H.~M.~Lee and P.~Sharma,
JHEP {\bf 1211}, 106 (2012);
A. Melfo, M. Nemev\v sek, F. Nesti, G. Senjanovi\'c and Y. Zhang,
Phys. Rev. D 85,  055018 (2012);
  M.~Aoki, S.~Kanemura and K.~Yagyu,
  Phys.\ Rev.\ D {\bf 85}, 055007 (2012);
  S.~Kanemura and K.~Yagyu,
  Phys.\ Rev.\ D {\bf 85}, 115009 (2012);
  M.~Aoki, S.~Kanemura, M.~Kikuchi and K.~Yagyu,
  [arXiv:1211.6029 [hep-ph]];
  E.~J.~Chun and P.~Sharma,
  Phys.\ Lett.\ B {\bf 722}, 86 (2013);
  L.~Wang and X.~-F.~Han,
  Phys.\ Rev.\ D {\bf 87}, 015015 (2013);
  P.~Dey, A.~Kundu and B.~Mukhopadhyaya,
  J.\ Phys.\ G {\bf 36}, 025002 (2009).
  
\bibitem{Hambye:2003ka} 
  T.~Hambye and G.~Senjanovic,
  Phys.\ Lett.\ B {\bf 582}, 73 (2004).

  
\bibitem{Ma:1998dx} 
  E.~Ma and U.~Sarkar,
  Phys.\ Rev.\ Lett.\  {\bf 80}, 5716 (1998);
  T.~Hambye, E.~Ma and U.~Sarkar,
  Nucl.\ Phys.\ B {\bf 602}, 23 (2001).
  
\bibitem{Chakraborty:2014xqa} 
  I.~Chakraborty and A.~Kundu,
  Phys.\ Rev.\ D {\bf 89}, 095032 (2014)
  
\bibitem{Chang:2014aaa} 
  X.~Chang and R.~Huo,
  Phys.\ Rev.\ D {\bf 89}, no. 3, 036005 (2014).
    
 \bibitem{FileviezPerez:2008bj}
  P.~Fileviez Perez, H.~H.~Patel, M.~.J.~Ramsey-Musolf and K.~Wang,
  Phys.\ Rev.\ D {\bf 79}, 055024 (2009);
  T.~Araki, C.~Q.~Geng and K.~I.~Nagao,
  Phys.\ Rev.\ D {\bf 83}, 075014 (2011);
  Y.~Kajiyama, H.~Okada and K.~Yagyu,
  Nucl.\ Phys.\ B {\bf 874}, 198 (2013):
  P.~V.~Dong, T.~P.~Nguyen and D.~V.~Soa,
  Phys.\ Rev.\ D {\bf 88}, 095014 (2013).
  

\bibitem{Kanemura:2012rj} 
  S.~Kanemura and H.~Sugiyama,
  Phys.\ Rev.\ D {\bf 86}, 073006 (2012).
  
\bibitem{Joglekar:2012vc}
  A.~Joglekar, P.~Schwaller and C.~E.~M.~Wagner,
  JHEP {\bf 1212} (2012) 064;
  M.~Fairbairn and P.~Grothaus,
  JHEP {\bf 1310}, 176 (2013).
  
\bibitem{Arina:2012aj} 
  C.~Arina, R.~N.~Mohapatra and N.~Sahu,
  Phys.\ Lett.\ B {\bf 720}, 130 (2013).
  
\bibitem{Dijkstra:2004cc} 
  T.~P.~T.~Dijkstra, L.~R.~Huiszoon and A.~N.~Schellekens,
  Nucl.\ Phys.\ B {\bf 710}, 3 (2005);
  O.~Lebedev, H.~P.~Nilles, S.~Raby, S.~Ramos-Sanchez, M.~Ratz, P.~K.~S.~Vaudrevange and A.~Wingerter,
  Phys.\ Lett.\ B {\bf 645}, 88 (2007).

  
\bibitem{Bahrami:2014ska} 
  S.~Bahrami and M.~Frank,
  Phys.\ Rev.\ D {\bf 90}, 035017 (2014);
  S.~Bahrami and M.~Frank,
  Phys.\ Rev.\ D {\bf 88}, 095002 (2013).
  
   
   
\bibitem{Keung:2011zc} 
  W.~Y.~Keung and P.~Schwaller,
  JHEP {\bf 1106}, 054 (2011).
  
  
\bibitem{Angle:2008we} 
  J.~Angle, E.~Aprile, F.~Arneodo, L.~Baudis, A.~Bernstein, A.~Bolozdynya, L.~C.~C.~Coelho and C.~E.~Dahl {\it et al.},
  Phys.\ Rev.\ Lett.\  {\bf 101}, 091301 (2008).
  
\bibitem{Ade:2013zuv} 
  P.~A.~R.~Ade {\it et al.}  [Planck Collaboration],
  Astron.\ Astrophys.\  {\bf 571}, A16 (2014).
  
  
\bibitem{Kanemura:2012rs} 
  S.~Kanemura and K.~Yagyu,
  Phys.\ Rev.\ D {\bf 85}, 115009 (2012);
  M.~Aoki, S.~Kanemura, M.~Kikuchi and K.~Yagyu,
  Phys.\ Rev.\ D {\bf 87}, 015012 (2013).
   
\bibitem{Arbabifar:2012bd} 
  F.~Arbabifar, S.~Bahrami and M.~Frank,
  Phys.\ Rev.\ D {\bf 87}, 015020 (2013).
 
 
\bibitem{Barate:2003sz} 
  R.~Barate {\it et al.}  [LEP Working Group for Higgs boson searches and ALEPH and DELPHI and L3 and OPAL Collaborations],
  Phys.\ Lett.\ B {\bf 565}, 61 (2003).

\bibitem{Ishiwata:2013gma} 
  K.~Ishiwata and M.~B.~Wise,
  arXiv:1307.1112 [hep-ph].
  
\bibitem{Lange:2014fxa} 
  C.~Lange [ATLAS for the and CMS Collaborations],
  arXiv:1411.7279 [hep-ex]; 
  G.~Aad {\it et al.}  [ATLAS Collaboration],
  Phys.\ Rev.\ Lett.\  {\bf 112}, 201802 (2014); 
  S.~Chatrchyan {\it et al.}  [CMS Collaboration],
  Eur.\ Phys.\ J.\ C {\bf 74}, no. 8, 2980 (2014);
  N.~Zhou, Z.~Khechadoorian, D.~Whiteson and T.~M.~P.~Tait,
  Phys.\ Rev.\ Lett.\  {\bf 113}, no. 15, 151801 (2014).
  
\bibitem{Belanger:2013xza} 
  G.~Belanger, B.~Dumont, U.~Ellwanger, J.~F.~Gunion and S.~Kraml,
  Phys.\ Rev.\ D {\bf 88}, 075008 (2013); 
  J.~R.~Espinosa, M.~Muhlleitner, C.~Grojean and M.~Trott,
  JHEP {\bf 1209}, 126 (2012)


	

\bibitem{Heikinheimo:2012yd} 
  M.~Heikinheimo, K.~Tuominen and J.~Virkajarvi,
  JHEP {\bf 1207}, 117 (2012).
  
  	
\bibitem{Komatsu:2010fb} 
  E.~Komatsu {\it et al.}  [WMAP Collaboration],
  Astrophys.\ J.\ Suppl.\  {\bf 192}, 18 (2011).
  
\bibitem{Belyaev:2012qa} 
  A.~Belyaev, N.~D.~Christensen and A.~Pukhov,
  Comput.\ Phys.\ Commun.\  {\bf 184}, 1729 (2013).
	
\bibitem{Belanger:2013oya} 
  G.~Belanger, F.~Boudjema, A.~Pukhov and A.~Semenov,
  Comput.\ Phys.\ Commun.\  {\bf 185}, 960 (2014).
  
\bibitem{Bernabei:2010mq} 
  R.~Bernabei {\it et al.}  [DAMA and LIBRA Collaborations],
  Eur.\ Phys.\ J.\ C {\bf 67}, 39 (2010).
  
\bibitem{Aalseth:2010vx} 
  C.~E.~Aalseth {\it et al.}  [CoGeNT Collaboration],
  Phys.\ Rev.\ Lett.\  {\bf 106}, 131301 (2011).
	
\bibitem{Angloher:2014myn} 
  G.~Angloher {\it et al.}  [CRESST-II Collaboration],
  arXiv:1407.3146 [astro-ph.CO].


\bibitem{Behnke:2012ys} 
  E.~Behnke {\it et al.}  [COUPP Collaboration],
  Phys.\ Rev.\ D {\bf 86}, no. 5, 052001 (2012)
  [Erratum-ibid.\ D {\bf 90}, no. 7, 079902 (2014)].

	
\bibitem{Aprile:2013doa} 
  E.~Aprile {\it et al.}  [XENON100 Collaboration],
  Phys.\ Rev.\ Lett.\  {\bf 111}, no. 2, 021301 (2013).
	
	
\bibitem{Lavina:2013zxa} 
  L.~S.~Lavina [ XENON100 Collaboration],
  arXiv:1305.0224 [hep-ex]; 
  E.~Aprile {\it et al.}  [XENON100 Collaboration],
  Phys.\ Rev.\ Lett.\  {\bf 109}, 181301 (2012);
  E.~Aprile {\it et al.}  [XENON100 Collaboration],
  Phys.\ Rev.\ Lett.\  {\bf 109}, 181301 (2012).

	
	


	
\bibitem{Agnese:2013cvt} 
  R.~Agnese {\it et al.}  [CDMS Collaboration],
  Phys.\ Rev.\ D {\bf 88}, 031104 (2013).
	
	
\bibitem{Li:2013fla} 
  H.~B.~Li {\it et al.}  [TEXONO Collaboration],
  Phys.\ Rev.\ Lett.\  {\bf 110}, no. 26, 261301 (2013).
	
\bibitem{Chavarria:2014ika} 
  A.~Chavarria, J.~Tiffenberg, A.~Aguilar-Arevalo, D.~Amidei, X.~Bertou, G.~Cancelo, J.~C.~D'Olivo and J.~Estrada {\it et al.},
  arXiv:1407.0347 [physics.ins-det].
	
	
\bibitem{Ackermann:2013yva} 
  M.~Ackermann {\it et al.}  [Fermi-LAT Collaboration],
  Phys.\ Rev.\ D {\bf 89}, no. 4, 042001 (2014).
  
\bibitem{Adriani:2008zr} 
  O.~Adriani {\it et al.}  [PAMELA Collaboration],
  Nature {\bf 458}, 607 (2009).
  
\bibitem{Accardo:2014lma} 
  L.~Accardo {\it et al.}  [AMS Collaboration],
  Phys.\ Rev.\ Lett.\  {\bf 113}, no. 12, 121101 (2014).
     
\bibitem{Aartsen:2013bka} 
  M.~G.~Aartsen {\it et al.}  [IceCube Collaboration],
  Phys.\ Rev.\ Lett.\  {\bf 111}, 021103 (2013).
  	
\bibitem{Avrorin:2014swy} 
  A.~D.~Avrorin {\it et al.}  [Baikal Collaboration],
  Astroparticle Physics (2015), pp. 12-20.
  
\bibitem{Boliev:2013ai} 
  M.~M.~Boliev, S.~V.~Demidov, S.~P.~Mikheyev and O.~V.~Suvorova,
  JCAP {\bf 1309}, 019 (2013).
  
\end{thebibliography}
\end{document}